\documentclass{JHEP3}

\usepackage{amsmath}
\usepackage{amsfonts}
\usepackage{amssymb}
\usepackage{amscd}
\usepackage{graphicx}
\usepackage{citesort}
\usepackage{mathrsfs}
\usepackage{bbm}
\usepackage{graphics}

\def\ad#1{{ #1}}

\def\S{{\mathbb S}}

\def\[{\begin{equation}}
\def\]{\end{equation}}
\def\<{\begin{eqnarray}}
\def\>{\end{eqnarray}}

\newcommand{\alg}[1]{\mathfrak{#1}}
\newcommand{\su}{\alg{su}}

\newcommand{\mathsym}[1]{{}}

\newcommand{\stateA}[1]{|#1\rangle^{\rm{I}}}
\newcommand{\stateB}[1]{|#1\rangle^{\rm{II}}}
\newcommand{\stateC}[1]{|#1\rangle^{\rm{III}}}

\def\defeq{\ \raisebox{0.2pt}{:}\hspace{-1.2mm}=}
\def\ds{\displaystyle}
\def\ssp{\hspace{0.3mm}}
\def\({\left(}
\def\){\right)}
\renewcommand{\[}{\left[}
\renewcommand{\]}{\right]}

\def\pare#1{\left\{ #1\right\}}
\def\vev#1{\langle #1\rangle}

\renewcommand{\eqref}[1]{$\({\rm \ref{#1}}\)$}

\usepackage{amsmath}
\usepackage{amsfonts}
\usepackage{amssymb}
\usepackage{amscd}
\usepackage{graphicx}
\usepackage{citesort}
\usepackage{mathrsfs}
\usepackage{bbm}

\def\S{{\mathbb S}}

\def\ads{{\rm AdS}_5\times {\rm S}^5}
\author{Gleb Arutyunov$^a$\footnote{ Correspondent fellow at
Steklov Mathematical Institute, Moscow.}, Marius de Leeuw$^a$, Ryo
Suzuki$^b$\,  and\,  Alessandro Torrielli$^a$\footnote{Emails:
G.E.Arutyunov@uu.nl, M.deLeeuw@uu.nl, rsuzuki@maths.tcd.ie,
A.Torrielli@uu.nl}
 \\
  \\ $^{a}$ {\it Institute for Theoretical
Physics and Spinoza Institute,\\ ~~Utrecht University, 3508 TD
Utrecht, The Netherlands} \\ \vskip 0.05cm $^b$ {\it School of
Mathematics, Trinity College, Dublin 2, Ireland}}

\title{Bound State Transfer Matrix for $\ads$ Superstring}

\abstract{We apply the algebraic Bethe ansatz technique to compute
the eigenvalues of the transfer matrix constructed from the
general bound state S-matrix of the light-cone $\ads$ superstring.
This allows us to verify certain conjectures on the quantum
characteristic function, and to extend them to the general case. }

\preprint{
          \tiny{ITP-UU-09-25}\\[-.5ex]
          \tiny{SPIN-09-24}\\[-.5ex]
          \tiny{TCDMATH 09-16}\\[-.5ex]
          }

\begin{document}

\section{Introduction}

The current challenge in the study of integrability in AdS/CFT
\cite{Minahan:2002ve,Kazakov:2004qf,Beisert:2004hm,Arutyunov:2004vx,Staudacher:2004tk,Beisert:2005fw,Beisert:2005tm,Janik:2006dc,Hofman:2006xt,Dorey:2006dq,Arutyunov:2006ak,Beisert:2006ib,Beisert:2006ez,Klose:2006zd,Arutyunov:2009ga}
is to find an exact solution to the planar spectral problem for
composite gauge-invariant operators of finite length. The success
obtained in the computation of L\"uscher's corrections to
anomalous dimensions in the framework of the string sigma model
\cite{Ambjorn:2005wa,Bajnok:2008bm,Fiamberti:2007rj,Bajnok:2008qj,Beccaria:2009eq,Beccaria:2009ny,Bajnok:2009vm},
combined with encouraging indications concerning the applicability
of the TBA approach
\cite{Arutyunov:2007tc,Arutyunov:2009zu,Gromov:2009tv,Bombardelli:2009ns,Gromov:2009bc,Arutyunov:2009ur,Arutyunov:2009kf,Frolov:2009in,Hegedus:2009ky,Correa:2009mz,Gromov:2009zb,Roiban:2009aa},
strengthens the expectations that such a solution may be reached
in the near future.

\smallskip

The various attempts appearing in the literature have partially been
based on powerful conjectures obtained by comparison
with well known integrable models, and by applying procedures that
are well established in standard cases. The confirmations obtained
through direct computation have then reinforced these conjectures.
However, the AdS/CFT integrable model is in itself very peculiar,
and its distinguished features make one wonder how far the
analogies with previously studied cases can be stretched.
Therefore, it is important to be able to provide an alternative
way to derive, from first principles, the relevant system of
equations characterizing the solution to the problem.

\smallskip

General criteria of integrability dictate that the complete
solution of the finite-size problem
can be obtained, once the full
set of (bound) states of the asymptotic spectrum, and all their
scattering matrices, are known\footnote{Strictly speaking, this is only known for the ground state.}.
Explicit fulfilment of this last
requirement was missing until very recently, when the complete set
of bound state S-matrices has been determined
\cite{Arutyunov:2009mi}. This opens the possibility to directly
evaluate the associated transfer matrix, a quantity which plays an
important role in the recent studies of the TBA and Y-system of
the AdS/CFT integrable model in \cite{Gromov:2009tv,Gromov:2009bc}
and also in \cite{Gromov:2009zb}. Finding eigenvalues of the
transfer matrix corresponding to generic bound state
representations is precisely the task we perform in this paper.

\smallskip

The method we will use consists in constructing the monodromy
matrix in arbitrary bound state representations, by using the
general expression for the S-matrix describing scattering of two
bound states, and to diagonalize the corresponding transfer matrix
by means of the Algebraic Bethe Ansatz (ABA) technique. As usual,
the spectrum is generated by applying certain operator entries of
the monodromy matrix to a chosen vacuum. Requiring the
corresponding state to be an eigenstate of the transfer matrix
results into the full set of eigenvalues and the associated Bethe
equations. This program has been already carried out for the
transfer matrix with all legs in the fundamental representation
\cite{Martins:2007hb}, and our result generalizes it to arbitrary
bound state representations.

\smallskip

Our general result is the formula (\ref{eqn;FullEignvalue}) for
spectrum of the transfer matrix, to be supplemented with the
auxiliary Bethe equations (\ref{bennote}). When restricting this general formula to
the $\alg{su}(2)$ vacuum (symmetric representation defined in the
main body of the paper), we obtain the following expression
\begin{eqnarray}
&&\Lambda(q|\vec{p}) = 1+\prod_{i=1}^{K^{\rm{I}}}
\left[\frac{(x^-_0-x^-_i)(1-x^-_0 x^+_i)}{(x^-_0-x^+_i)(1-x^+_0
x^+_i)}\sqrt{\frac{x^+_0x^+_i}{x^-_0x^-_i}}\mathscr{X}^{\ell_0,0}_{\ell_0}\right]\nonumber\\
\nonumber
&&~~~~-2\sum_{k=0}^{\ell_0-1}\prod_{i=1}^{K^{\rm{I}}}\left(\frac{x^+_0-x^+_i}{x^-_0-x^+_i}
\sqrt{\frac{x^-_0}{x^+_0}}\left[1-\frac{k}{u_0-u_i+\frac{\ell_0-\ell_i}{2}
}\right]\mathscr{X}^{k,0}_k\right) +
\sum_{a=\pm}\sum_{k=1}^{\ell_0-1}\prod_{i=1}^{K^{\rm{I}}}\lambda_a(q,p_i,k)\,
,
%+\sum_{k=1}^{\ell_0-1}
%\prod_{i=1}^{K^{\rm{I}}}\lambda_-(q,p_i,k).\nonumber
\end{eqnarray}
where $\mathscr{X}^{\ell_0,0}_{\ell_0}$ is given by (\ref{xloloo}), $\mathscr{X}^{k,0}_{k}$ by (\ref{xlolook}) and $\lambda_\pm$ by (\ref{eqn;lambda-pm}).
In the case of fundamental representations in the physical space,
we can also make a different choice of the vacuum, and select the
$\alg{sl}(2)$ vacuum (which corresponds to antisymmetric
representations). The formula we obtain in this case is
(\ref{anti}), which we also present here for convenience:

\begin{eqnarray*}
\Lambda(q|\vec{p})=&&(\ell_0+1)\prod_{i=1}^{K^{\rm{I}}}\frac{x_0^--x^-_i}{x_0^+-x^-_i}
\sqrt{\frac{x^+_0}{x^-_0}}-
\ell_0\prod_{i=1}^{K^{\rm{I}}}\frac{x_0^--x_i^+}{x^+_0-x^-_i}
\sqrt{\frac{x^+_0x^-_i}{x^-_0x^+_i}}\, -\\\nonumber &&-\ell_0
\prod_{i=1}^{K^{\rm{I}}}\frac{x_0^--x_i^-}{x^+_0-x^-_i}\frac{x_i^--\frac{1}{x_0^+}}
{x_i^+-\frac{1}{x_0^+}}\sqrt{\frac{x^+_0x^+_i}{x^-_0x^-_i}}
+(\ell_0-1)\prod_{i=1}^{K^{\rm{I}}}\frac{x_0^--x^+_i}{x_0^+-x^-_i}
\frac{x_i^--\frac{1}{x_0^+}}{x_i^+-\frac{1}{x_0^+}}\sqrt{\frac{x^+_0}{x^-_0}}.\nonumber
\end{eqnarray*}

\smallskip

After  obtaining the explicit solution, we are in the position of
comparing with certain results and conjectures existing in the
literature. First of all, we re-obtain the set of bound state
Bethe equations derived in \cite{deLeeuw:2008ye}. Secondly, by
studying the eigenvalues reported above, with all the physical legs in the fundamental representation,
we were able to confirm the expression for the transfer matrix
proposed by \cite{Beisert:2006qh}, and utilized in
\cite{Gromov:2009tv}. This agreement is found both in the
symmetric and in the antisymmetric representation. In
\cite{Beisert:2006qh}, the analysis was based on fusion properties
and the expansion of a quantum characteristic function. This is an
operator used for the construction of transfer matrix eigenvalues
in various symmetric representations, in the context of Baxter and
Hirota equations \cite{Krichever:1996qd}. An educated guess for
the present situation was made in \cite{Beisert:2006qh},
generalizing the proposal in \cite{Beisert:2005di}. See also the
papers \cite{Tsuboi:1997iq,Kazakov:2007fy,Kazakov:2007na}, where
the Bazhanov-Reshetikhin \cite{Bazhanov:1989yk} formula for fusion
has been considered in the case of standard $\alg{gl}(m|n)$
superalgebras. Our formulas therefore simultaneously furnish a
proof of these conjectured properties in the centrally-extended
case, and provide a generalization to arbitrary bound states.

Most importantly, the transfer matrix in the symmetric
representation obtained through the fusion procedure
\cite{Beisert:2006qh} exhibits an explicit dependence on the
kinematical parameters corresponding to the bound state
constituents. The result we find shows that this dependence is, in
fact, artificial, since we are able to unambiguously re-expresses
the corresponding eigenvalue purely in terms of parameters
characterizing the bound state as a whole.

As a check, we have also reproduced the L\"uscher corrections
found in the literature
\cite{Bajnok:2008bm,Bajnok:2008qj,Gromov:2009tv} directly from the
transfer matrix eigenvalues. The results we obtain allow one to
enlarge the set of operators for which L\"uscher's corrections can
be explicitly found to those which correspond to bound states of
the world-sheet theory.

We point out that we have explicitly carried out the full
calculation of the transfer-matrix eigenvalues corresponding only
to the $\alg{su}(2)$ and $\alg{sl}(2)$ vacua, to one excitation
over the $\alg{su}(2)$ vacuum\footnote{The double-excitation case
already involves a number of terms of order one hundred, and its
complete treatment would therefore involve a much bigger effort.},
and to a certain class of excited states (see appendix
\ref{sec;TasBA}). This, together with the fact that the formulas
we deduce perfectly agree with those obtained {\it via} the
expansion of the quantum characteristic function (even with an
arbitrary configuration of auxiliary roots) makes us confident
about our findings.

\smallskip

The paper is organized as follows. In the next section we recall
the necessary facts concerning the construction of the bound state
scattering matrix and define the corresponding transfer matrix. In
section 3 we use the ABA technique to diagonalize this transfer
matrix. We discuss two particular choices of the vacuum,
corresponding to the highest weight state of the totally symmetric
and anti-symmetric representations of $\su(2|2)$. Finally, we report in several
appendices useful material and computations.

\section{Scattering and transfer matrices of the string sigma model}

In what follows we will apply the algebraic Bethe ansatz technique
to diagonalize the transfer matrix corresponding to bound state
representations of the $\ads$ string sigma model. The emerging
Bethe ansatz exhibits  a nested structure and, therefore,  to keep
the discussion transparent, in this section we will fix and
explain the notation used throughout the paper. We will also
recall some relevant facts about the bound state representations.

\subsection{Bound states and their S-matrix}

In the uniform light-cone gauge \cite{Arutyunov:2004yx} the
symmetry algebra of the $\ads$ string sigma model in the
decompactification limit is (two copies of) the centrally extended
$\alg{su}(2|2)$ superalgebra \cite{Arutyunov:2006ak}. The latter
is also a symmetry algebra of the spin chain Hamiltonian
associated with $\mathcal{N}=4$ super Yang-Mills theory
\cite{Beisert:2005tm}. The asymptotic spectrum of the light-cone
sigma model consists of elementary particles, {\it i.e.} the ones
transforming in the fundamental representations of
$\alg{su}(2|2)$, and of their bound states. An $\ell$-particle
bound state transforms in the tensor product of two
$4\ell$-dimensional atypical totally symmetric multiplets of
$\alg{su}(2|2)$.

The centrally-extended algebra $\alg{su}(2|2)$ consists of bosonic generators
$\mathbb{R},\mathbb{L}$ (generating two copies of
$\mathfrak{su}(2)$), supersymmetry generators
$\mathbb{Q},\mathbb{G}$ and central charges
$\mathbb{H},\mathbb{C},\mathbb{C}^{\dag}$.
The generators $\mathbb{Q},\mathbb{G}$ are conjugate to each
other. For the string sigma model, $\mathbb{H}$ corresponds to the
light-cone Hamiltonian and the central charge $\mathbb{C}$ is a
function of the world-sheet momentum.

A convenient way to deal with bound state representations and the
S-matrix action is to use the superspace formalism
\cite{Arutyunov:2008zt}. One considers the vector space of
analytic functions $\Phi(w,\theta)$ of two bosonic variables
$w_{1,2}$ and two fermionic variables $\theta_{3,4}$. Any such
function can be expanded as
\begin{eqnarray}
\Phi(w,\theta) &=&\sum_{\ell=0}^{\infty}\Phi_{\ell}(w,\theta),\nonumber\\
\Phi_{\ell} &=& \phi^{a_{1}\ldots a_{\ell}}w_{a_{1}}\ldots
w_{a_{\ell}} +\phi^{a_{1}\ldots a_{\ell-1}\alpha}w_{a_{1}}\ldots
w_{a_{\ell-1}}\theta_{\alpha}+\nonumber\\
&&\phi^{a_{1}\ldots a_{\ell-2}34} \, w_{a_{1}}\ldots
w_{a_{\ell-2}}\theta_{3}\theta_{4}.
\end{eqnarray}
Restriction to $\Phi_{\ell}$ furnishes an atypical totally
symmetric representation of dimension $4\ell$.  It is realized on
a graded vector space with basis $|e_{a_{1}\ldots
a_{\ell}}\rangle, |e_{a_{1}\ldots
a_{\ell-1}\alpha}\rangle,|e_{a_{1}\ldots a_{\ell-2}34}\rangle$,
where $a_{i}$ are bosonic indices and $\alpha,\beta$ are fermionic
indices, and each of the basis vectors is totally symmetric in the
bosonic indices and anti-symmetric in the fermionic indices. In
terms of the above analytic functions, the basis vectors of the
totally symmetric representation can be identified as
$|e_{a_{1}\ldots a_{\ell}}\rangle \leftrightarrow w_{a_{1}}\ldots
w_{a_{\ell}},|e_{a_{1}\ldots a_{\ell-1}\alpha}\rangle
\leftrightarrow w_{a_{1}}\ldots w_{a_{\ell-1}}\theta_{\alpha}$ and
$|e_{a_{1}\ldots a_{\ell-1}34}\rangle \leftrightarrow
w_{a_{1}}\ldots w_{a_{\ell-2}}\theta_{3}\theta_{4}$, respectively.
In the superspace formalism, the algebra generators are
represented by differential operators.

Consider two-particle states. We denote the bound state numbers of
the scattering particles as $\ell_1$ and $\ell_2$, respectively.
In superspace the tensor product of the corresponding bound state
representations  is given by
\begin{eqnarray}
\Phi_{\ell_1}(w,\theta)\Phi_{\ell_2}(v,\vartheta),
\end{eqnarray}
where $w,\theta$ denote the superspace variables of the first
particle and $v,\vartheta$ describe the representation of the
second particle.

Because of ${\alg{su}(2)}\times {\alg{su}(2)}$ invariance, when
acting on such a tensor product representation space, the
S-matrix\footnote{In this picture the S-matrix is understood as an
operator $\S : V_1 \otimes V_2 \longrightarrow V_1 \otimes V_2$.}
$\S$ leaves  invariant five different subspaces
\cite{Arutyunov:2009mi}. Each of these subspaces is characterized
by a specific assignment of ${\alg{su}(2)}\times {\alg{su}(2)}$
Dynkin labels, which are quantum numbers that are trivially
conserved under scattering. Two pairs of these subspaces are
simply related to each other by exchanging the type of fermions
appearing, as described below. It leaves only three non-equivalent
cases, which we list here below \cite{Arutyunov:2009mi}.

\subsection*{Case I}
The standard basis for this vector space, which we will concisely
call $V^{\rm{I}}$, is
\begin{eqnarray}\label{eqn;BasisCase1}
\stateA{k,l}\equiv\underbrace{\theta_{3}w_1^{\ell_1-k-1}w_2^{k}}_{\rm{Space
1}}~\underbrace{\vartheta_{3}v_1^{\ell_2-l-1}v_2^{l}}_{\rm{Space
2}},
\end{eqnarray}
for all $k+l=N$. The range of $k,l$ here and in the cases below is
straightforwardly read off from the definition of the states.
In particular, $k$ ranges from $0$ to $\ell_1-1$, and $l$ ranges
from $0$ to $\ell_2-1$. For fixed $N$, this gives in this case
$N+1$ different vectors. We get another copy of Case I if we
exchange the index $3$ with $4$ in the fermionic variable, with
the same S-matrix.

\subsection*{Case II}

The standard basis for this space $V^{\rm{II}}$ is
\begin{eqnarray}\label{eqn;BasisCase2}
\stateB{k,l}_1&\equiv& \underbrace{\theta_{3}w_1^{\ell_1-k-1}w_2^{k}}~\underbrace{v_1^{\ell_2-l}v_2^{l}},\nonumber\\
\stateB{k,l}_2&\equiv&\underbrace{w_1^{\ell_1-k}w_2^{k}}~\underbrace{\vartheta_{3}v_1^{\ell_2-l-1}v_2^{l}},\\
\stateB{k,l}_3&\equiv&\underbrace{\theta_{3}w_1^{\ell_1-k-1}w_2^{k}}~\underbrace{\vartheta_{3}\vartheta_{4}v_1^{\ell_2-l-1}v_2^{l-1}},\nonumber\\
\stateB{k,l}_4&\equiv&\underbrace{\theta_{3}\theta_{4}w_1^{\ell_1-k-1}w_2^{k-1}}~\underbrace{\vartheta_{3}v_1^{\ell_2-l-1}v_2^{l}},\nonumber
\end{eqnarray}
where $k+l=N$ as before\footnote{We will from now on, with no risk
of confusion, omit indicating ``Space 1" and ``Space 2" under the
curly brackets.}. It is easily seen that we get in this case
$4N+2$ states. Once again, exchanging $3$ with $4$ in the
fermionic variable gives another copy of Case II, with the same
S-matrix.

\subsection*{Case III}

For fixed $N=k+l$, the dimension of this vector space
$V^{\rm{III}}$ is $6N$. The standard basis is
\begin{eqnarray}\label{eqn;BasisCase3}
\stateC{k,l}_1&\equiv&\underbrace{w_1^{\ell_1-k}w_2^{k}}~\underbrace{v_1^{\ell_2-l}v_2^{l}},\nonumber\\
\stateC{k,l}_2&\equiv&\underbrace{w_1^{\ell_1-k}w_2^{k}}~\underbrace{\vartheta_{3}\vartheta_{4}v_1^{\ell_2-l-1}v_2^{l-1}},\nonumber\\
\stateC{k,l}_3&\equiv&\underbrace{\theta_{3}\theta_{4}w_1^{\ell_1-k-1}w_2^{k-1}}~\underbrace{v_1^{\ell_2-l}v_2^{l}},\nonumber\\
\stateC{k,l}_4&\equiv&\underbrace{\theta_{3}\theta_{4}w_1^{\ell_1-k-1}w_2^{k-1}}~\underbrace{\vartheta_{3}\vartheta_{4}v_1^{\ell_2-l-1}v_2^{l-1}},\\
\stateC{k,l}_5&\equiv&\underbrace{\theta_{3}w_1^{\ell_1-k-1}w_2^{k}}~\underbrace{\vartheta_{4}v_1^{\ell_2-l}v_2^{l-1}},\nonumber\\
\stateC{k,l}_6&\equiv&\underbrace{\theta_{4}w_1^{\ell_1-k}w_2^{k-1}}~\underbrace{\vartheta_{3}v_1^{\ell_2-l-1}v_2^{l}}\nonumber.
\end{eqnarray}

As was explained in \cite{Arutyunov:2009mi}, the different cases
are mapped into one another by use of the (opposite) coproducts of
the (Yangian) symmetry generators. The S-matrix has the
following block-diagonal form
\begin{eqnarray}
\S=\begin{pmatrix}
  \fbox{\small{$\mathscr{X}$}} & ~ & ~ & ~ & ~ \\
  ~ & \fbox{\LARGE{$\mathscr{Y}$}} & ~ & \mbox{\Huge{$0$}} & ~ \\
  ~ & ~ & \fbox{\Huge{$\mathscr{Z}$}} & ~ & ~ \\
  ~ & \mbox{\Huge{$0$}} & ~ & \fbox{\LARGE{$\mathscr{Y}$}} & ~ \\
  ~ & ~ & ~ & ~ & \fbox{\small{$\mathscr{X}$}}
\end{pmatrix}.
\end{eqnarray}
The outer blocks scatter states from $V^{\rm{I}}$
\begin{eqnarray}
&&\mathscr{X}:V^{\rm{I}}\longrightarrow V^{\rm{I}},\\
&&\stateA{k,l}\mapsto \sum_{m=0}^{k+l}
\mathscr{X}^{k,l}_m\stateA{m,k+l-m},
\end{eqnarray}
where the explicit form of the coefficients $\mathscr{X}^{k,l}_n$
is given in appendix \ref{A}. The blocks $\mathscr{Y}$ describe
the scattering of states from $V^{\rm{II}}$
\begin{eqnarray}
&&\mathscr{Y}:V^{\rm{II}}\longrightarrow V^{\rm{II}},\\
&&\stateB{k,l}_i\mapsto \sum_{m=0}^{k+l}\sum_{j=1}^{4}
\mathscr{Y}^{k,l;j}_{m;i}\stateB{m,k+l-m}_j.
\end{eqnarray}
These S-matrix elements are given in Eq. (5.18) of
\cite{Arutyunov:2009mi}, but we will not need their general
expression here. We will only need some special cases, which we
have listed in appendix \ref{A}. Finally, the middle block deals
with the third case
\begin{eqnarray}
&&\mathscr{Z}:V^{\rm{III}}\longrightarrow V^{\rm{III}},\\
&&\stateC{k,l}_i\mapsto \sum_{m=0}^{k+l}\sum_{j=1}^{6}
\mathscr{Z}^{k,l;j}_{m;i}\stateC{m,k+l-m}_j,
\end{eqnarray}
with $\mathscr{Z}^{k,l;j}_{m;i}$ from Eq. (6.11) in
\cite{Arutyunov:2009mi}. Similarly, their general expression is
not needed for the sake of the present discussion, and we refer to
\cite{Arutyunov:2009mi} for details. Once again, the few ones that
we will actually need are also listed in appendix \ref{A}.

The S-matrix above is canonically normalized\footnote{ This agrees
for example with the normalization of the matrix part of
\cite{Arutyunov:2009mi} and \cite{Bajnok:2008bm}.}, {\it i.e.}, on
the vector $|0,0\rangle^{\rm III}_1=w_1^{\ell_1}v_1^{\ell_2}$, the
action of the S-matrix is $\S|0,0\rangle^{\rm
III}_1=|0,0\rangle^{\rm III}_1$. The full $\ads$ string bound
state S-matrix, in the ${\alg{su}}(2)$ sector, is then obtained by
taking two copies of the above S-matrix, and multiplying the
result by a scalar factor. The latter is determined through the
fusion procedure by using the scalar factor of the fundamental
S-matrix  \cite{Chen:2006gq,Arutyunov:2008zt}.

\subsection{Monodromy and transfer matrices}

Having introduced the relevant vector spaces and S-matrix, we are
ready to define the corresponding monodromy and transfer matrix.

Consider $K^{\rm{I}}$ bound state particles with bound state
numbers $\ell_1,\ldots,\ell_{K^{\rm{I}}}$ and momenta
$p_1,\ldots,p_{K^{\rm{I}}}$. To these particles we add an
auxiliary one, with momentum $q$ and bound state number $\ell_0$.
Any state of this system lives in the following tensor
product space
\begin{eqnarray}
\mathcal{V}:=V_{\ell_0}(q)\otimes V_{\ell_1}(p_1)\otimes \ldots
\otimes V_{\ell_{K^{\rm{I}}}}(p_{K^{\rm{I}}}),
\end{eqnarray}
where $V_{\ell_i}$ is a carrier space of the bound state
representation with the number $\ell_i$. We split the states in
this space into an auxiliary piece and a physical piece:
\begin{eqnarray}
|A\rangle_0\otimes|B\rangle_{K^{\rm{I}}} \in \mathcal{V},\nonumber
\end{eqnarray}
where $ |A\rangle_0 \in V_{\ell_0}(q)$ and\footnote{All the \ad{tensor}
products are defined with increasing order of the index as
$1,2,\ldots, K$.} $|B\rangle_{K^{\rm{I}}}
 \in  V_P:=\bigotimes_{i} V_{\ell_i}(p_{i})$. The monodromy matrix
acting in the space $\mathcal{V}$ is defined as follows
\begin{eqnarray}
\mathcal{T}_{\ell_0}(q|\vec{p}) :=
\prod_{i=1}^{K^{\rm{I}}}\mathbb{S}_{0k}(q,p_i),
\end{eqnarray}
where $\mathbb{S}_{0k}(q,p_k)$ is the bound state S-matrix
describing scattering between the auxiliary particle and a
`physical' particle with momentum $p_k$ and bound state number
$\ell_k$.

 The monodromy matrix can be seen as a $4\ell_0\times 4\ell_0$
dimensional matrix in the auxiliary space $V_{\ell_0}(q)$,  the
corresponding matrix elements  being themselves operators on
$V_P$. Indeed, introducing a basis $|e_{I}\rangle$ for
$V_{\ell_0}(q)$, with the index $I$ labelling a $4 \ell_0$-dimensional space, and a basis
$|f_{A}\rangle$ for $V_P$, the action of the monodromy matrix
$\mathcal{T}\equiv \mathcal{T}_{\ell_0}(q|\vec{p})$ on the total
space $\mathcal{V}$ can be written as
\begin{eqnarray}\label{eqn;ActionMonodromy}
\mathcal{T}(|e_{I}\rangle\otimes |f_A\rangle) = \sum_{J,B}
T_{IA}^{JB}|e_{J}\rangle\otimes |f_B\rangle.
\end{eqnarray}
The matrix entries of the monodromy matrix can then be denoted as
\begin{eqnarray}
\mathcal{T}|e_{I}\rangle = \sum_{J} \mathcal{T}^{J}_{I}
|e_{J}\rangle\,  ,
\end{eqnarray}
while the action of the matrix elements $\mathcal{T}^{J}_{I}$ as
operators on $V_P$ can easily be read off:
\begin{eqnarray}
\mathcal{T}^{J}_{I}|f_A\rangle = \sum_{B} T_{IA}^{JB}|f_B\rangle.
\end{eqnarray}
The operators $\mathcal{T}^{J}_{I}$ have non-trivial commutation
relations among themselves. Consider two different auxiliary
spaces $V_{\ell_0}(q),V_{\tilde{\ell}_0}(\tilde{q})$. The
Yang-Baxter equation for $\S$ implies that
\begin{eqnarray}\label{eqn;YBE-operators}
\mathbb{S}(q,\tilde{q})\mathcal{T}_{\ell_0}(q|\vec{p})\mathcal{T}_{\tilde{\ell}_0}(\tilde{q}|\vec{p})
=\mathcal{T}_{\tilde{\ell}_0}(\tilde{q}|\vec{p})\mathcal{T}_{\ell_0}(q|\vec{p})\mathbb{S}(q,\tilde{q}),
\end{eqnarray}
where $\mathbb{S}(q,\tilde{q})$ is the S-matrix describing the
scattering between two bound state particles of bound state
numbers $\ell_0,\tilde{\ell}_0$ and momenta $q,\tilde{q}$
respectively. By explicitly working out these relations, one finds
the commutation relations between the different matrix elements of
the monodromy matrix. The fundamental commutation relations
(\ref{eqn;YBE-operators}) constitute a cornerstone of the
Algebraic Bethe Ansatz \cite{Faddeev:1996iy}.

It is convenient to pick up the following explicit basis
$|e_{I}\rangle$ in the space $V_{\ell_0}(q)$
\begin{equation}
\label{basisV0}
\begin{aligned}
e_{\alpha;k} &:= \theta_{\alpha} w_1^{\ell_0-k-1}w_{2}^{k},\\
e_{k} &:= w_1^{\ell_0-k}w_{2}^{k},\\
e_{34;k} &:= \theta_{3}\theta_{4} w_1^{\ell_0-k-1}w_{2}^{k-1}.
\end{aligned}
\end{equation}
The transfer matrix is then defined as
\begin{eqnarray}
\mathscr{T}_0(q|\vec{p}):={\rm{str}}_0\mathcal{T}_{\ell_0}(q|\vec{p}),
\end{eqnarray}
and it can be viewed as an operator acting on the physical space
$V_P$. In terms of the operator entries of the monodromy matrix,
the transfer matrix is written as
\begin{eqnarray}\label{eqn;Transfer}
\mathscr{T}_0(q|\vec{p})=\sum_{k=0}^{\ell_0} \mathcal{T}^{k}_{k} + \sum_{k=1}^{\ell_0-1} \mathcal{T}^{34;k}_{34;k} -
\sum_{k=0}^{\ell_0-1} \, \sum_{\alpha=3,4} \mathcal{T}^{\alpha;k}_{\alpha;k}.
\end{eqnarray}
In the remainder of this paper we will study the action of
$\mathscr{T}_0(q|\vec{p})$ on the physical space in detail and derive its eigenvalues.

\section{Diagonalization of the transfer matrix}

An efficient way to find the eigenvalues of the transfer matrix is
to use the Algebraic Bethe Ansatz. We start by defining a vacuum
state
\begin{eqnarray}
|0\rangle_P = w_1^{\ell_1}\otimes\ldots\otimes w_1^{\ell_{K^{\rm{I}}}}.
\end{eqnarray}
We then compute the action of the transfer matrix on this state,
which appears to be one of its eigenstates, and afterwards use specific
elements of the monodromy matrix to generate the whole spectrum of
eigenvalues. Imposing the eigenstate condition should result in the
determination of the full set of eigenvalues and associated Bethe
equations, therefore providing the complete solution of the
asymptotic spectral problem.

\subsection{Eigenvalue of the transfer matrix on the vacuum}
As promised, we first deduce the action of the transfer matrix on
the vacuum. We will do this for each of the separate sums in
(\ref{eqn;Transfer}). Let us start with the fermionic part, {\it
i.e.} we want to compute
\begin{eqnarray}
\sum_{k=0}^{\ell_0-1}
\mathcal{T}^{\alpha;k}_{\alpha;k}|0\rangle_P, \qquad \alpha=3,4.
\end{eqnarray}
Taking into account the explicit form of the S-matrix elements
entering the monodromy matrix, we find that the only contribution
to $\mathcal{T}^{\alpha;k}_{\alpha;k}|0\rangle$ comes from
diagonal scattering elements. To be precise, one finds
\begin{eqnarray}
\mathcal{T}^{\alpha;k}_{\alpha;k}|0\rangle_P=\prod_{i}\mathscr{Y}^{k,0;1}_{k;1}(q,p_i)|0\rangle_P,
\end{eqnarray}
where $\mathscr{Y}^{k,0;1}_{k;1}(q,p_i)$ are Case II S-matrix
elements. By explicitly working out this expression, one finds
\begin{eqnarray}
\mathscr{Y}^{k,0;1}_{k;1}(q,p_i) &=&
\frac{x^+_0-x^+_i}{x^-_0-x^+_i}\sqrt{\frac{x^-_0}{x^+_0}}\left[1-\frac{k}{u_0-u_i+\frac{\ell_0-\ell_i}{2}
}\right]\mathscr{X}^{k,0}_k (q, p_i),
\end{eqnarray}
where $x^{\pm}_0$ are defined in terms of the momentum $q$, and
one uses equation (4.13) of \cite{Arutyunov:2009mi}:
\begin{alignat}{3}
\mathscr{X}^{k,0}_{k} (q,p_i) &=
\mathcal{D} \, \frac{\prod_{j=0}^{k - 1} u_0 - u_i + \frac{\ell_0 - \ell_i - 2j}{2}}{\prod_{j=1}^{k} u_0 - u_i + \frac{\ell_0 + \ell_i - 2j}{2}} \ad{\qquad k =1,\cdots,\ell_0 -1}, \notag \\[1mm]
\ad{\mathscr{X}^{0,0}_0 (q,p_i)} &\ad{= \mathcal{D} \,
\frac{\pi}{\sin (\pi \ell_i) \Gamma (1-\ell_i) \Gamma(\ell_i)} = \mathcal{D} = \frac{x_0^- - x_i^+}{x_0^+ - x_i^-} \sqrt{\frac{x_0^+}{x_0^-} \, \frac{x_i^-}{x_i^+} }\,.}
\label{xlolook}
\end{alignat}
Obviously, the contribution of $\mathcal{T}^{\alpha;k}_{\alpha;k}$
is the same for $\alpha=3,4$. Here $x_m$, $m=0,1,...K^I$, are the
constrained parameters ($\lambda$ is the 't Hooft coupling)
$$
x_m^++\frac{1}{x^+_m}-x_m^--\frac{1}{x^-_m}=2\ell_m\frac{i}{g}\, ,
~~~~~g=\frac{\sqrt{\lambda}}{2\pi}
$$
related to the particle momenta as
$p_m=\frac{1}{i}\log\Big(\frac{x^+_m}{x^-_m}\Big)$. Also, $u_m$
represents the corresponding rapidity variable given by
\begin{eqnarray}
\label{uvar} x_m^{\pm}+\frac{1}{x^{\pm}_m}=\frac{2 i}{g}u_m\pm
\frac{i}{g}\ell_m\, .
\end{eqnarray}
Next, we consider the more involved  bosonic part. This can be
written as
\begin{eqnarray}
\mathcal{T}^{0}_{0} + \mathcal{T}^{\ell_0}_{\ell_0} + \sum_{k=1}^{\ell_0-1}\left\{\mathcal{T}^{k}_{k} + \mathcal{T}^{34;k}_{34;k}\right\}.
\end{eqnarray}
We first determine $\mathcal{T}^{0}_{0}|0\rangle_P$ and
$\mathcal{T}^{\ell_0}_{\ell_0}|0\rangle_P$. For these operators,
one again finds that only diagonal scattering elements of the
S-matrices contribute, which leads to
\begin{eqnarray}
\begin{aligned}
\mathcal{T}^{0}_{0}|0\rangle_P &=\prod_{i} \mathscr{Z}^{0,0;1}_{0;1}(q,p_i)  \, |0\rangle_P,\\
\mathcal{T}^{\ell_0}_{\ell_0}|0\rangle_P&= \prod_{i}
\mathscr{Z}^{\ell_0,0;1}_{\ell_0;1}(q,p_i) |0\rangle_P.
\end{aligned}
\end{eqnarray}
These matrix elements can be computed explicitly and give
\begin{eqnarray}
\mathcal{T}^{0}_{0}|0\rangle_P &=&|0\rangle_P,\\
\mathcal{T}^{\ell_0}_{\ell_0}|0\rangle_P&=&
\left\{\prod_{i=1}^{K^{\rm{I}}} \frac{(x^-_0-x^-_i)(1-x^-_0
x^+_i)}{(x^-_0-x^+_i)(1-x^+_0
x^+_i)}\sqrt{\frac{x^+_0x^+_i}{x^-_0x^-_i}}\mathscr{X}^{\ell_0,0}_{\ell_0}(q,p_i)\right\} |0\rangle_P \ad{\,.}
\end{eqnarray}
\ad{where we define
\begin{equation}
\label{xloloo}
\mathscr{X}^{\ell_0,0}_{\ell_0} (q,p_i)=
\mathcal{D} \, \frac{\prod_{j=0}^{\ell_0 - 1} u_0 - u_i + \frac{\ell_0 - \ell_i - 2j}{2}}{\prod_{j=1}^{\ell_0} u_0 - u_i + \frac{\ell_0 + \ell_i - 2j}{2}} \,.
\end{equation}}
The next thing to consider is the sum
\begin{eqnarray}
\sum_{k=1}^{\ell_0-1}\left\{\mathcal{T}^{k}_{k} + \mathcal{T}^{34;k}_{34;k}\right\}.
\end{eqnarray}
While in the previous computations one could simply restrict to
the diagonal elements, one obtains instead a matrix structure for
this last piece. This is due to the fact that there are scattering
processes that relate $w_2 \leftrightarrow
\theta_{\alpha}\theta_{\beta}$. To be more precise, for the action
of $\mathcal{T}^k_k$ and $\mathcal{T}^{34,k}_{34,k}$ one finds
\begin{eqnarray}
\mathcal{T}^{k}_{k}|0\rangle_P &=&\sum_{a_1\ldots
a_{K^{\rm{I}}}=1,3}\mathscr{Z}^{k,0;a_1}_{k;1}(q,p_1) \, \mathscr{Z}^{k,0;a_2}_{k;a_1}(q,p_2)\ldots\mathscr{Z}^{k,0;1}_{k;a_{K^{\rm{I}}}}(q,p_{K^{\rm{I}}})|0\rangle_P,\\
\mathcal{T}^{34,k}_{34,k}|0\rangle_P&=&
\sum_{a_1\ldots
a_{K^{\rm{I}}}=1,3}\mathscr{Z}^{k,0;a_1}_{k;3}(q,p_1)\, \mathscr{Z}^{k,0;a_2}_{k;a_1}(q,p_2)\ldots\mathscr{Z}^{k,0;3}_{k;a_{K^{\rm{I}}}}(q,p_{K^{\rm{I}}})|0\rangle_P.
\end{eqnarray}
In order to evaluate the above expressions explicitly, it proves
useful to use a slightly more general reformulation\footnote{We remark that this computation has been performed at weak coupling in \cite{Bajnok:2008bm,Bajnok:2008qj}.}. One can reintroduce the
elements $\mathcal{T}^{34,k}_{k}$ and $\mathcal{T}^{k}_{
34,k}$ from the monodromy matrix. Their action on the
vacuum is
\begin{eqnarray}
\mathcal{T}^{34,k}_{k}|0\rangle_P &=&\sum_{a_1\ldots
a_{K^{\rm{I}}}=1,3}\mathscr{Z}^{k,0;a_1}_{k;1}(q,p_1)\, \mathscr{Z}^{k,0;a_2}_{k;a_1}(q,p_2)\ldots\mathscr{Z}^{k,0;3}_{k;a_{K^{\rm{I}}}}(q,p_{K^{\rm{I}}})|0\rangle_P,\\
\mathcal{T}^{k}_{34,k}|0\rangle_P&=& \sum_{a_1\ldots
a_{K^{\rm{I}}}=1,3}\mathscr{Z}^{k,0;a_1}_{k;3}(q,p_1)\, \mathscr{Z}^{k,0;a_2}_{k;a_1}(q,p_2)\ldots\mathscr{Z}^{k,0;1}_{k;a_{K^{\rm{I}}}}(q,p_{K^{\rm{I}}})|0\rangle_P.
\end{eqnarray}
They describe the mixing between the states
$|e_{34,k}\rangle$ and $|e_{k}\rangle$. If we consider
the two-dimensional vector space spanned by
$|e_{34,k}\rangle$ and $|e_{k}\rangle$ for fixed $k \in \{1,...,\ell_0 -1\}$, we
see that the above elements define a $2\times2$ dimensional matrix
\begin{eqnarray}
\mathcal{T}_{2\times2}=
\begin{pmatrix}
    \mathcal{T}^{k}_{k} & \mathcal{T}^{34,k}_{k}\\
    \mathcal{T}^{k}_{34,k} & \mathcal{T}^{34,k}_{34,k}
\end{pmatrix},
\end{eqnarray}
and the bosonic part of the transfer matrix is just the trace of
this matrix. Moreover, it is easily seen from the definition of
the transfer matrix that this matrix factorizes
\begin{eqnarray}
\mathcal{T}_{2\times2} = \prod_{i=1}^{K}
\begin{pmatrix}
    \mathscr{Z}^{k,0;1}_{k;1}(q,p_i) & \mathscr{Z}^{k,0;3}_{k;1}(q,p_i)\\
    \mathscr{Z}^{k,0;1}_{k;3}(q,p_i) & \mathscr{Z}^{k,0;3}_{k;3}(q,p_i)
\end{pmatrix}.
\end{eqnarray}
The trace of this matrix is given by the sum of its eigenvalues,
hence it remains to find the eigenvalues of this matrix.
Actually, it is easily checked that the eigenvectors of
\begin{eqnarray}
\begin{pmatrix}
    \mathscr{Z}^{k,0;1}_{k;1}(q,p_i) & \mathscr{Z}^{k,0;3}_{k;1}(q,p_i)\\
    \mathscr{Z}^{k,0;1}_{k;3}(q,p_i) & \mathscr{Z}^{k,0;3}_{k;3}(q,p_i)
\end{pmatrix}
\end{eqnarray}
are independent of $p_i$. In other words, these are automatically
eigenvectors of $\mathcal{T}_{2\times2}$, and the corresponding eigenvalues are
the product of the eigenvalues of the above matrices. The
individual eigenvalues are given by
\begin{eqnarray}\label{eqn;lambda-pm}
\lambda_\pm(q,p_i,k)&=&\frac{\mathscr{X}^{k,0}_k}{2\mathcal{D}}\left[1-\frac{(x^-_ix^+_0-1)
   (x^+_0-x^+_i)}{(x^-_i-x^+_0)
   (x^+_0x^+_i-1)}+\frac{2ik}{g}\frac{x^+_0
   (x^-_i+x^+_i)}{(x^-_i-x^+_0)
   (x^+_0x^+_i-1)}\right.\\
&&\left.\pm\frac{i x^+_0
   (x^-_i-x^+_i)}{(x^-_i-x^+_0)
 (x^+_0x^+_i-1)}\sqrt{\left(\frac{2k}{g}\right)^2+2i
\left[x^+_0+\frac{1}{x^+_0}\right]
\frac{2k}{g}-\left[x^+_0-\frac{1}{x^+_0}\right]^2}\right]\nonumber.
\end{eqnarray}
The action of the transfer matrix on the vacuum is now given by
the sum of all the above terms. From this it is easily seen that
the vacuum is indeed an eigenvector of the transfer matrix with
the following eigenvalue
\begin{eqnarray}\label{transfer-rankone}
\Lambda(q|\vec{p})&=&1+\prod_{i=1}^{K^{\rm{I}}}
\left[\frac{(x^-_0-x^-_i)(1-x^-_0 x^+_i)}{(x^-_0-x^+_i)(1-x^+_0
x^+_i)}\sqrt{\frac{x^+_0x^+_i}{x^-_0x^-_i}}\mathscr{X}^{\ell_0,0}_{\ell_0}\right]\nonumber\\
&&-2\sum_{k=0}^{\ell_0-1}\prod_{i=1}^{K^{\rm{I}}}\left(\frac{x^+_0-x^+_i}{x^-_0-x^+_i}
\sqrt{\frac{x^-_0}{x^+_0}}\left[1-\frac{k}{u_0-u_i+\frac{\ell_0-\ell_i}{2}
}\right]\mathscr{X}^{k,0}_k\right) \nonumber\\
&& +
\sum_{k=1}^{\ell_0-1}\prod_{i=1}^{K^{\rm{I}}}\lambda_+(q,p_i,k)
+\sum_{k=1}^{\ell_0-1} \prod_{i=1}^{K^{\rm{I}}}\lambda_-(q,p_i,k).
\end{eqnarray}
For the fundamental case ($\ell_0=\ell_i=1 \, \, \forall i$), this reduces to
\begin{eqnarray}\label{T fundamental-all}\nonumber
\mathcal{T}_0(q|\vec{p})|0\rangle_P
&=& \left\{\prod_{i}\mathscr{Z}^{0,0;1}_{0;1}(q,p_i) +
\prod_{i}\mathscr{Z}^{1,0;1}_{1;1}(q,p_i) -2
\prod_{i}\mathscr{Y}^{0,0;1}_{0;1}(q,p_i)\right\}|0\rangle_P\\
&=&\left\{1 + \prod_{i=1}^{K^{\rm{I}}}
\frac{1-\frac{1}{x^-_0x^+_i}}{1-\frac{1}{x^-_0x^-_i}}\frac{x^+_0-x^+_i}{x^+_0-x^-_i}
-2
\prod_{i=1}^{K^{\rm{I}}}\frac{x^+_0-x^+_i}{x^+_0-x^-_i}\sqrt{\frac{x^-_i}{x^+_i}}\right\}|0\rangle_P.
\end{eqnarray}

We would like to point out that the square roots
in the eigenvalues $\lambda_{\pm}$ will never appear in the vacuum
eigenvalue. This is because the square root part only depends on
the auxiliary momentum $q$, and it can be seen that, after summing
the contribution from $\lambda_+$ and $\lambda_-$, only even powers
of this square root piece survive.

\subsection{Creation operators and excited states}
The next step in the algebraic Bethe ansatz is to introduce
creation operators. These operators will be entries from our
monodromy matrix. By acting with those operators on the vacuum one
creates new (excited) states, which again will be eigenstates of
the transfer matrix. We will need to specify which monodromy
matrix entries correspond to creation operators for our purposes.

Recall that, from the symmetry invariance of the S-matrix, one can
deduce that the quantum numbers $K^{\rm{II}}$ (total number of
fermions) and $K^{\rm{III}}$ (total number of fermions of a
definite species, say, $3$) are conserved. Any element
$\mathcal{T}^{J}_{I}$ is called a creation operator if
$K^{\rm{II}}(|e_I\rangle_0)>K^{\rm{II}}(|e_J\rangle_0)$, it is
called an annihilation operator if
$K^{\rm{II}}(|e_I\rangle_0)<K^{\rm{II}}(|e_J\rangle_0)$ and
diagonal if
$K^{\rm{II}}(|e_I\rangle_0)=K^{\rm{II}}(|e_J\rangle_0)$. The
reason for this assignment is the following. Consider a creation
operator $\mathcal{T}^{J}_{I}$ and any physical state
$|A\rangle_P$. The action of a creation operator is defined via
(\ref{eqn;ActionMonodromy}). Since the total number $K^{\rm{II}}$
is preserved, and the $K^{\rm{II}}$ charge in the auxiliary space
has decreased, it has necessarily increased in the physical space.
The number $K^{\rm{II}}$ corresponds to the number of fermions in
the system, hence, by acting with $\mathcal{T}^{J}_{I}$ on
$|A\rangle_P$, one creates extra fermions in the physical space.
Notice that this also implies that acting with an annihilation
operator on the vacuum annihilates it, whence the name.

We will create excited states by considering fundamental auxiliary
spaces with momenta $\lambda_i$. Since these are fundamental
spaces, their monodromy matrices are only $4\times4$-dimensional.
Our discussion will be very similar to the treatment of the
algebraic Bethe ansatz for the Hubbard model which was first
performed in \cite{martins-1997,Ramos:1996us}. In order to make
contact with the treatment of \cite{Martins:2007hb} and with the
standard notation used for the Hubbard model, we parameterize this
monodromy matrix as
\begin{eqnarray}
\begin{pmatrix}
B & B_{3} & B_{4} & F\\
C_{3} & A^{3}_{3} & A^{3}_{4} & B^{*}_{3}\\
C_{4} & A^{4}_{3} & A^{4}_{4} & B^{*}_{4}\\
C & C^{*}_{3} & C^{*}_{4} & D\\
\end{pmatrix}.
\end{eqnarray}
Notice that one finds two seemingly different sets of creation
operators
$B_{3}(\lambda_i),B_{4}(\lambda_i),F(\lambda_i)$ and
$B^*_{3}(\lambda_i),B^*_{4}(\lambda_i),F(\lambda_i)$. As
discussed in \cite{martins-1997}, it is enough to restrict to one
set. In what follows, we will use the operators
$B_{3}(\lambda_i),B_{4}(\lambda_i),F(\lambda_i)$ to
create fermionic excitations out of the vacuum.

A generic excited state will now be formed by acting with a number
of those operators on the vacuum, e.g. one can consider states
like
\begin{eqnarray}
B_{3}(\lambda_1)B_{4}(\lambda_2)|0\rangle.
\end{eqnarray}
To find out whether this is an eigenstate of the transfer matrix,
one has to commute the diagonal elements of the
transfer matrices through the creation operators and let them act
on the vacuum. Imposing the eigenstate condition will in general give constraints on the
momenta $\lambda_i$. The explicit commutation relations will be
the subject of the next section.

\subsection{Commutation relations}

In order to compute the action of the transfer matrix on an
excited state, we need to compute the commutation relations between
the diagonal elements $\mathcal{T}^A_A$ and the aforementioned
creation operators. While we have to use creation operators in a fundamental auxiliary representation, the diagonal
elements are to be taken in the bound state representation with generic $\ell_0$. The commutation relations follow from
(\ref{eqn;YBE-operators}). We will report the complete derivation of one specific commutation
relation, and only give the final result for the remaining ones. In the derivation, one has to
pay particular attention to the fermionic nature of the operators.

Consider the
operator $B_{3}(\lambda)$ and the element
$\mathcal{T}^{3,k}_{3,k}$ from the transfer matrix. From
(\ref{eqn;YBE-operators}), one finds
\begin{eqnarray}
\mathbb{P}_{3,k|0}\mathbb{S}(q,\lambda)\mathcal{T}(q)\mathcal{T}(\lambda)e_{3,k}\tilde{e}_{3,0}
=
\mathbb{P}_{3,k|0}\mathcal{T}(\lambda)\mathcal{T}(q)\mathbb{S}(q,\lambda)e_{3,k}\tilde{e}_{3,0},
\end{eqnarray}
where we have dropped the indices $\ell_0$ and $\tilde{\ell}_0=1$, and
where the tilde on $\tilde{e}_{3,0}$ denotes a basis element in the second auxiliary space.
The operator $\mathbb{P}_{A|B}$ is the projection operator onto the subspace generated by the basis element
$e_A\tilde{e}_B$. The right hand side of the above equation gives
\begin{eqnarray}
\mathbb{P}_{3,k|0}\mathcal{T}(\lambda)\mathcal{T}(q)\mathbb{S}(q,\lambda)e_{3,k}\tilde{e}_{3,0}&=&
\mathbb{P}_{3,k|0}\mathscr{X}^{k,0}_k\mathcal{T}(\lambda)\mathcal{T}(q)e_{3,k}\tilde{e}_{3,0}\nonumber\\
&=&\mathbb{P}_{3,k|0}\sum_{A,B}\mathscr{X}^{k,0}_k(-1)^{F_A}(\mathcal{T}_{3}^B\tilde{e}_B)(\lambda)(\mathcal{T}_{3,k}^{A}(q)e_{A})\nonumber\\
&=&\mathscr{X}^{k,0}_k(-1)^{F_{(3,k)}}(\mathcal{T}_{3}^0\tilde{e}_0)(\lambda)(\mathcal{T}_{3,k}^{3,k}(q)e_{3,k})\nonumber\\
&=&-\mathscr{X}^{k,0}_k
B_{3}^0(\lambda)\mathcal{T}_{3,k}^{3,k}(q)e_{3,k}\tilde{e}_0.
\end{eqnarray}
The left hand side reduces to
\begin{eqnarray}
\mathbb{P}_{3,k|0}\mathbb{S}(q,\lambda)\mathcal{T}(q)\mathcal{T}(\lambda)e_{3,k}\tilde{e}_{3,0}&=&
-\mathbb{P}_{3,k|0}\mathbb{S}(q,\lambda)\mathcal{T}(q)(\mathcal{T}(\lambda)_{3}^B\tilde{e}_{B})e_{3,k}\\
&=&-\mathbb{P}_{3,k|0}\mathbb{S}(q,\lambda)\mathcal{T}(q)
\left\{\mathcal{T}_{3}^0(\lambda)\tilde{e}_{0}+\mathcal{T}_{3}^{3}(\lambda)\tilde{e}_{3}+\mathcal{T}_{3}^1(\lambda)\tilde{e}_{1}\right\}e_{3,k}\nonumber\\
&=&\mathbb{P}_{3,k|0}\mathbb{S}(q,\lambda)(\mathcal{T}_{3,k}^A(q)e_A)
\left\{\mathcal{T}_{3}^0(\lambda)\tilde{e}_{0}+\mathcal{T}_{3}^{3}(\lambda)\tilde{e}_{3}+\mathcal{T}_{3}^1(\lambda)\tilde{e}_{1}\right\}.\nonumber
\end{eqnarray}
Because of the projection, we only need to take into account terms
that are mapped onto $e_{3,k}\tilde{e}_0$ by the action of the S-matrix. These are given
by
\begin{eqnarray}
&&\mathbb{P}_{3,k|0}\mathbb{S}(q,\lambda)\left\{\mathcal{T}_{3,k}^{3,k}(q)e_{3,k}\mathcal{T}_{3}^0(\lambda)\tilde{e}_{0}
+
\mathcal{T}_{3,k}^{3,k-1}(q)e_{3,k-1}\mathcal{T}_{3}^1(\lambda)\tilde{e}_{1}
+\mathcal{T}_{3,k}^{k}(q)e_{k}\mathcal{T}_{3}^{3}(\lambda)\tilde{e}_{3,0}\right\}=\nonumber\\
&&\qquad\mathbb{P}_{3,k|0}\mathbb{S}(q,\lambda)\left\{-\mathcal{T}_{3,k}^{3,k}(q)\mathcal{T}_{3}^0(\lambda)e_{3,k}\tilde{e}_{0}
-
\mathcal{T}_{3,k}^{3,k-1}(q)\mathcal{T}_{3}^1(\lambda)e_{3,k-1}\tilde{e}_{1}
+ \right.\nonumber\\
&&\qquad\qquad\left.+\mathcal{T}_{3,k}^{k}(q)\mathcal{T}_{3}^{3}(\lambda)e_{k}\tilde{e}_{3,0}+\mathcal{T}_{3,k}^{34,k-1}(q)\mathcal{T}_{3}^{3}(\lambda)e_{34,k-1}\tilde{e}_{3,0}\right\}.~~~~
\end{eqnarray}
Working this out explicitly yields
\begin{eqnarray}
&&\mathbb{P}_{3,k|0}\mathbb{S}(q,\lambda)\left\{\mathcal{T}_{3,k}^{3,k}(q)e_{3,k}\mathcal{T}_{3}^0(\lambda)\tilde{e}_{0}
+
\mathcal{T}_{3,k}^{3,k-1}(q)e_{3,k-1}\mathcal{T}_{3}^1(\lambda)\tilde{e}_{1}
+\mathcal{T}_{3,k}^{k}(q)e_{k}\mathcal{T}_{3}^{3}(\lambda)\tilde{e}_{3,0}\right\}=\nonumber\\
&&\qquad\left\{-\mathcal{T}_{3,k}^{3,k}(q)\mathcal{T}_{3}^0(\lambda)\mathscr{Y}^{k,0;1}_{k;1}-
\mathscr{Y}^{k-1,1;1}_{k;1}\mathcal{T}_{3,k}^{3,k-1}(q)\mathcal{T}_{3}^1(\lambda)
+\right.\nonumber\\
&&\qquad\qquad\left.+\mathscr{Y}^{k,1;1}_{k;2}\mathcal{T}_{3,k}^{k}(q)\mathcal{T}_{3}^{3}(\lambda)
+\mathscr{Y}^{k,1;1}_{k;4}\mathcal{T}_{3,k}^{34,k-1}(q)\mathcal{T}_{3}^{3}(\lambda)
\right\}e_{3,k}\tilde{e}_{0}.
\end{eqnarray}
From this we now read off the final commutation relation\footnote{Throughout the rest of this section 3.3, if not otherwise indicated, the coefficient functions appearing have to be understood as $\mathscr{X} \equiv \mathscr{X}(q,\lambda)$, $\mathscr{Y} \equiv \mathscr{Y}(q,\lambda)$, $\mathscr{Z} \equiv \mathscr{Z}(q,\lambda)$ (indices are omitted here for simplicity).}
\begin{eqnarray}\label{eqn;CommRel}
\mathscr{X}^{k,0}_k B_{3}(\lambda)\mathcal{T}_{3,k}^{3,k}(q)
&=&\mathscr{Y}^{k,0;1}_{k;1}\mathcal{T}_{3,k}^{3,k}(q)B_{3}(\lambda)+\mathscr{Y}^{k-1,1;1}_{k;1}\mathcal{T}_{3,k}^{3,k-1}(q)C^*_{3}(\lambda)+\\
&&-\mathscr{Y}^{k,1;1}_{k;2}\mathcal{T}_{3,k}^{k}(q)A_{3}^{3}(\lambda)-\mathscr{Y}^{k,1;1}_{k;4}\mathcal{T}_{3,k}^{34,k-1}(q)A_{3}^{3}(\lambda)\nonumber.
\end{eqnarray}
Notice that in the above relation the operators are ordered in
such a way that all annihilation and diagonal elements are on the
right. This is done because the action of
those elements on the vacuum is known. We would also like to compare these commutation relations
with \cite{martins-1997,Ramos:1996us} for the Hubbard model. We
see that the first and third term are also present in the Hubbard
model. However, due to the fact that we are dealing with bound
state representation, we also obtain {\it two additional terms}.

Generically, the commutation relations produce ``wanted" terms,
which are those which directly contribute to the eigenvalue, and other ``unwanted"
terms. The latter terms are those which need to vanish, in order for the state of our ansatz to be an
eigenstate. In (\ref{eqn;CommRel}), one can easily see by acting on the vacuum that the wanted term is the first
term on the right hand side, while the other terms are unwanted. The
cancellation of the unwanted terms will give rise to certain constraints, which are
precisely the auxiliary Bethe equations.

The other commutation relations one needs to compute are those
with $\mathcal{T}^{k}_{k},\mathcal{T}^{34,k}_{34,k}$ and
$\mathcal{T}^{4,k}_{4,k}$. Their derivation is considerably more involved,
especially the procedure of reordering them according to the above ``annihilation and diagonal on the right" prescription. We will present
the commutation relations we will actually need
in the coming sections. We will give the wanted terms, and focus on
one specific type of unwanted terms. Schematically, we will focus on the following structure:
\begin{eqnarray}\label{eqn;CommRel2}
\left[\mathcal{T}^k_k(q)
+\mathcal{T}^{\alpha\beta,k}_{\alpha\beta,k}(q)
\right]B_{\alpha}(\lambda) &=&
\frac{\mathscr{X}^{k,0}_k}{\mathscr{Y}^{k,0;1}_{k;1}}B_{\alpha}(\lambda)\left[\mathcal{T}^k_k(q)
+\mathcal{T}^{\alpha\beta,k}_{\alpha\beta,k}(q) \right] +
\frac{\mathscr{Y}^{k,0;1}_{k;2}}{\mathscr{Y}^{k,0;1}_{k;1}}\mathcal{T}^{k}_{\alpha,k}(q)B(\lambda)
+ \ldots\nonumber\\
\mathcal{T}^{\gamma,k}_{\alpha_1,k}(q)B_{\alpha_2}(\lambda) &=&
\frac{\mathscr{X}^{k,0}_k}{\mathscr{Y}^{k,0;1}_{k;1}}B_{\beta_2}(\lambda)\mathcal{T}^{\gamma,k}_{\beta_1,k}(q)
r_{\alpha_1\alpha_2}^{\beta_1\beta_2}(u_0+{\textstyle{\frac{\ell_0-1}{2}}}-k,u_{\lambda}) + \\
&&+
\frac{\mathscr{Y}^{k,0;1}_{k;2}}{\mathscr{Y}^{k,0;1}_{k;1}}\mathcal{T}^{k}_{\beta_1,k}(q)A^{\gamma}_{\beta_2}(\lambda)
r_{\alpha_1\alpha_2}^{\beta_1\beta_2}(u_{\lambda},u_{\lambda})+\nonumber\ldots
\end{eqnarray}
Here, $u_\lambda$ is given by (cf. (\ref{uvar}))

\begin{eqnarray}
u_\lambda \, = \, \frac{g}{2 i} \bigg( x^+ (\lambda) +
\frac{1}{x^+ (\lambda)} - \frac{i}{g} \bigg)\nonumber
\end{eqnarray}
and $r^{\gamma\delta}_{\alpha\beta}(u_\lambda,u_\mu)$ are the
components of the 6-vertex model S-matrix
(\ref{eqn;6vertexComponents}) with $U=-1$.
We would like to point out that, when comparing this structure
against formulas (34-36) of \cite{martins-1997}, one immediately recognizes a
similarity between the commutation relations. As was shown in
\cite{Martins:2007hb}, for the case in which all representations
are taken to be fundamental, the commutation relations do agree. The
additional contributions coming from the fact that we are dealing
with bound states in e.g. (\ref{eqn;CommRel}), will only generate a new
class of unwanted terms. Hence, these new terms will not contribute to the
eigenvalues.

Let us mention one commutation relation which is particularly straightforward to derive, namely, the one between two
fermionic creation operators, as found from
(\ref{eqn;YBE-operators}) with $\ell_0 = \tilde{\ell}_0 =1$. This relation reads
\begin{eqnarray}\label{eqn;CommRelCreation}
B_{\alpha}(\lambda)B_{\beta}(\mu) &=& -\mathscr{X}^{0,0}_0 (\lambda,\mu) \, B_{\delta}(\mu)B_{\gamma}(\lambda)r^{\gamma\delta}_{\alpha\beta}(u_\lambda,u_\mu)\nonumber\\
&&\qquad \qquad + \, \frac{\mathscr{Z}^{1,0;1}_{1;6}(\lambda,\mu)}{\mathscr{Z}^{1,0;1}_{1;1}(\lambda,\mu)}\left[F(\lambda)B(\mu)-F(\mu)B(\lambda)\right]\epsilon_{\alpha\beta}.\quad
~
\end{eqnarray}
This reproduces the
result of \cite{Martins:2007hb}, and, in this way, one can see the
emergence of nesting. As a matter of fact, in
\cite{Ramos:1996us,martins-1997,Martins:2007hb} the appearance of
the 6-vertex model S-matrix was used to completely fix
the form of the excited eigenstates, and this can also be
done in our case.

\subsection{First excited state}

The first excited state is of the form
\begin{eqnarray}
|1\rangle = \mathcal{F}^{\alpha}B_{\alpha}(\lambda)|0\rangle_P,
\end{eqnarray}
where we sum over the repeated fermionic index. This state has
$K^{\rm{II}}=1$. As previously discussed, all the
commutation relations are ordered in such a way that all
annihilation and diagonal operators are on the right. From the
commutation relations (\ref{eqn;CommRel2}) one finds
\begin{eqnarray}
&&\left[\mathcal{T}^k_k(q)
+\mathcal{T}^{\alpha\beta,k}_{\alpha\beta,k}(q) \right]
\mathcal{F}^{\alpha}B_{\alpha}(\lambda)|0\rangle_P =
\frac{\mathscr{X}^{k,0}_k}{\mathscr{Y}^{k,0;1}_{k;1}}
\mathcal{F}^{\alpha}B_{\alpha}(\lambda)\left[\mathcal{T}^k_k(q)
+\mathcal{T}^{\alpha\beta,k}_{\alpha\beta,k}(q)
\right]|0\rangle_P\nonumber\\
&&\qquad \qquad \qquad \qquad \qquad \qquad \qquad \qquad +\frac{\mathscr{Y}^{k,0;1}_{k;2}}{\mathscr{Y}^{k,0;1}_{k;1}} \mathcal{F}^{\alpha}\mathcal{T}^{k}_{\alpha,k}(q)B(\lambda)|0\rangle_P,\\
&&\left[\mathcal{T}^{\alpha_1,k}_{\alpha_1,k}(q)\right]
\mathcal{F}^{\alpha_2}B_{\alpha_2}(\lambda)|0\rangle_P =
\frac{\mathscr{X}^{k,0}_k}{\mathscr{Y}^{k,0;1}_{k;1}}
\mathcal{F}^{\alpha_2}r_{\alpha_1\alpha_2}^{\beta_1\beta_2}(u_0+{\textstyle{\frac{\ell_0-1}{2}}}-k,u_{\lambda})B_{\beta_2}(\lambda)\left[\mathcal{T}^{\beta_1,k}_{\alpha_1,k}(q)\right]
|0\rangle_P\nonumber \\
&& \qquad \qquad \qquad \qquad \qquad \qquad +
\frac{\mathscr{Y}^{k,0;1}_{k;2}}{\mathscr{Y}^{k,0;1}_{k;1}}\mathcal{F}^{\alpha_2}r_{\alpha_1\alpha_2}^{\beta_1\beta_2}(u_{\lambda},u_{\lambda})\mathcal{T}^{k}_{\beta_2,k}(q)A^{\alpha_1}_{\beta_1}(\lambda)
|0\rangle_P,
\end{eqnarray}
where we remind that we concentrate on only one type of
unwanted terms, for the sake of clarity. The coefficient functions appearing in the above two formulas have to be understood as
$\mathscr{X} \equiv \mathscr{X}(q,\lambda)$, $\mathscr{Y} = \mathscr{Y}(q,\lambda)$ (indices are omitted here for simplicity).

Since $\mathcal{T}^{\alpha,k}_{\beta,k}|0\rangle_P \sim
\delta^{\alpha}_{\beta}|0\rangle_P $, we find that $|1\rangle$ can
only be an eigenstate of the transfer matrix if
\begin{eqnarray}
\mathcal{F}^{\alpha}r_{\gamma\alpha}^{\gamma\beta}(u_0+{\textstyle{\frac{\ell_0-1}{2}}}-k,u_{\lambda})
\sim \mathcal{F}^{\beta}.
\end{eqnarray}
This means that $\mathcal{F}^{\alpha}$ is an eigenvector of the
transfer matrix of the 6-vertex model. Luckily, one finds that the
eigenstates of the 6-vertex model are independent of the auxiliary
momenta. This means that the $k$ dependence in the above r-matrix
only appears in the eigenvalue $\Lambda^{(6v)}$, where $\Lambda^{(6v)}$ is the eigenvalue of the auxiliary 6-vertex
model. From
(\ref{eqn;eigenvalue6V}) we find ($K=K^{\rm{II}}=1$)
\begin{eqnarray}
\Lambda^{(6v)}(u_0 |u_\lambda ) &=& \prod_{i=1}^{K^{\rm{III}}}
\frac{1}{b(w_i,u_0+{\textstyle{\frac{\ell_0-1}{2}}}-k)}+b(u_0+{\textstyle{\frac{\ell_0-1}{2}}}-k,u_\lambda )\prod_{i=1}^{K^{\rm{III}}}
\frac{1}{b(u_0+{\textstyle{\frac{\ell_0-1}{2}}}-k,w_i)} ,\quad \nonumber
\end{eqnarray}
together with the auxiliary equation (\ref{eqn;AuxEqns6V})
\begin{eqnarray}
b(w_j,u_\lambda )=\prod_{i=1,i\neq j}^{K^{\rm{III}}}
\frac{b(w_j,w_i)}{b(w_i,w_j)}.
\end{eqnarray}
We also have to deal with the unwanted terms. Here we remark that,
since we have chosen $\mathcal{F}^{\alpha}$ to be an eigenvector
of the 6-vertex S-matrix, this also affects the unwanted terms.
One explicitly finds that they are proportional to
\begin{eqnarray}
\left\{\Lambda^{(6v)}(u_\lambda |u_\lambda )A_{\alpha}^{\alpha}(\lambda)-B(\lambda)\right\}|0\rangle_P.
\end{eqnarray}
Explicitly working this out this leads us to the following
auxiliary Bethe equations:
\begin{eqnarray}
\prod_{i=1}^{K^{\rm{I}}}\frac{x^+(\lambda)-x^-_i}{x^+(\lambda)-x^+_i}\sqrt{\frac{x^+_i}{x^-_i}}=\Lambda^{(6v)}(u_\lambda |u_\lambda ).
\end{eqnarray}
In order to make contact with the bound state Bethe equations
found in \cite{deLeeuw:2008ye}, let us define $y\equiv
x^+(\lambda)$ and rescale $w \to \frac{g}{2i}w$. We find that
$|1\rangle$ is an eigenstate, provided the auxiliary Bethe
equations hold\footnote{We remark that, for $K^{\rm{III}}=0$, the
solution of (\ref{bete}) correspond to the highest weight state of
the auxiliary six-vertex model, while, for $K^{\rm{III}}=1$, one
formally obtains a solution only if some of the auxiliary roots
are equal to infinity. This corresponds to a descendent of the
highest weight state under the $\alg{su}(2)$ symmetry.}:
\begin{eqnarray}
\label{bete}
\prod_{i=1}^{K^{\rm{I}}}\frac{y-x^-_i}{y-x^+_i}\sqrt{\frac{x^+_i}{x^-_i}}&=&
\prod_{i=1}^{K^{\rm{III}}}\frac{w_i-y-\frac{1}{y}-\frac{i}{g}}{w_i-y-\frac{1}{y}+\frac{i}{g}},
\nonumber\\
\frac{w_i-y-\frac{1}{y}+\frac{i}{g}}{w_i-y-\frac{1}{y}-\frac{i}{g}}
&=& \prod_{j=1,j\neq
i}^{K^{\rm{III}}}\frac{w_i-w_j+\frac{2i}{g}}{w_i-w_j-\frac{2i}{g}}.
\end{eqnarray}
This exactly matches with the auxiliary bound state Bethe ansatz
equations. The corresponding eigenvalue is
\begin{eqnarray}\label{eqn;Lambda1exc}
&&\Lambda(q|\vec{p})={\textstyle{\frac{y-x^-_0}{y-x^+_0}\sqrt{\frac{x^+_0}{x^-_0}}
+}}\\
&&
{\textstyle{+}}{\textstyle{\frac{y-x^-_0}{y-x^+_0}\sqrt{\frac{x^+_0}{x^-_0}}
\left[\frac{x^+_0+\frac{1}{x^+_0}-y-\frac{1}{y}}{x^+_0+\frac{1}{x^+_0}-y-\frac{1}{y}-\frac{2i\ell_0}{g}}\right]}}
\prod_{i=1}^{K^{\rm{I}}}
{\textstyle{\left[\frac{(x^-_0-x^-_i)(1-x^-_0
x^+_i)}{(x^-_0-x^+_i)(1-x^+_0
x^+_i)}\sqrt{\frac{x^+_0x^+_i}{x^-_0x^-_i}}\mathscr{X}^{\ell_0,0}_{\ell_0}\right]}}\nonumber\\
&&{\textstyle{+}}
\sum_{k=1}^{\ell_0-1}{\textstyle{\frac{y-x^-_0}{y-x^+_0}\sqrt{\frac{x^+_0}{x^-_0}}
\left[\frac{x^+_0+\frac{1}{x^+_0}-y-\frac{1}{y}}{x^+_0+\frac{1}{x^+_0}-y-\frac{1}{y}-\frac{2ik}{g}}\right]}}
\left\{\prod_{i=1}^{K^{\rm{I}}}{\textstyle{\lambda_+(q,p_i,k)+}}\right.\left.\prod_{i=1}^{K^{\rm{I}}}{\textstyle{\lambda_-(q,p_i,k)}}\right\}\nonumber\\
&&\quad
{\textstyle{-}}\sum_{k=0}^{\ell_0-1}{\textstyle{\frac{y-x^-_0}{y-x^+_0}\sqrt{\frac{x^+_0}{x^-_0}}
\left[\frac{x^+_0+\frac{1}{x^+_0}-y-\frac{1}{y}}{x^+_0+\frac{1}{x^+_0}-y-\frac{1}{y}-\frac{2ik}{g}}\right]}}
\prod_{i=1}^{K^{\rm{I}}}{\textstyle{\frac{x^+_0-x^+_i}{x^-_0-x^+_i}\sqrt{\frac{x^-_0}{x^+_0}}\left[1-\frac{k}{u_0-u_i+\frac{\ell_0-\ell_i}{2}
}\right]}}\times\nonumber\\
&&\quad\times
{\textstyle{\mathscr{X}^{k,0}_k}}\left\{\prod_{i=1}^{K^{\rm{III}}}{\textstyle{\frac{w_i-x^+_0-\frac{1}{x^+_0}+\frac{i(2k-1)}{g}}{w_i-x^+_0-\frac{1}{x^+_0}+\frac{i(2k+1)}{g}}+
}}{\textstyle{\frac{y+\frac{1}{y}-x^+_0-\frac{1}{x^+_0}+\frac{2ik}{g}}{y+\frac{1}{y}-x^+_0-\frac{1}{x^+_0}+\frac{2i(k+1)}{g}}}}\prod_{i=1}^{K^{\rm{III}}}{\textstyle{\frac{w_i-x^+_0-\frac{1}{x^+_0}+\frac{i(2k+3)}{g}}{w_i-x^+_0-\frac{1}{x^+_0}+\frac{i(2k+1)}{g}}}}\right\}.\nonumber
\end{eqnarray}
We stress once again that the above eigenvalue is for the canonically normalized
S-matrix. The dependence of $\Lambda$ on the bound state numbers
of the physical particles is hidden in their parameters
$x_i^{\pm}$ and in the S-matrix element $\mathscr{X}$. Notice that, when projected in the fundamental representation,
the formula above reproduces
the result of \cite{Martins:2007hb}.

\subsection{General result and Bethe equations}

As was stressed before, by comparing our commutation relations
against (34)-(36) from \cite{martins-1997,Ramos:1996us}, one
immediately notices several similarities. It turns out that one
can closely follow the derivation presented in those papers, and
from the diagonal terms read off the general eigenvalue.
Furthermore, cancelling the first few unwanted terms reveals
itself as sufficient to derive the complete set of auxiliary Bethe
equations.

More specifically, the results of Appendix \ref{sec;TasBA} and the
previously known results for the case when all physical legs are in the fundamental representation indicate
the generalization of the formula for the transfer-matrix eigenvalues to multiple excitations. In terms of S-matrix
elements, this generalization is given by
\begin{eqnarray}
\Lambda(q|\vec{p}) &=&
\prod_{m=1}^{K^{\rm{II}}}\frac{\mathscr{X}^{0,0}_0(q,\lambda_m)}{\mathscr{Y}^{0,0;1}_{0;1}(q,\lambda_m)}
+ \prod_{i=1}^{K^{\rm{I}}} \mathscr{Z}^{\ell_0,0;1}_{\ell_0;1}(q,p_i)\prod_{m=1}^{K^{\rm{II}}}\frac{\mathscr{X}^{\ell_0,0}_{\ell_0}(q,\lambda_m)}{\mathscr{Y}^{\ell_0,0;1}_{\ell_0;1}(q,\lambda_m)}+\nonumber\\
&&\sum_{k=1}^{\ell_0-1}
\prod_{m=1}^{K^{\rm{II}}}\frac{\mathscr{X}^{k,0}_k(q,\lambda_m)}{\mathscr{Y}^{k,0;1}_{k;1}(q,\lambda_m)}\left\{\prod_{i=1}^{K^{\rm{I}}}\lambda_+(q,p_i)
+\prod_{i=1}^{K^{\rm{I}}}\lambda_-(q,p_i)\right\}+\nonumber\\
&&-\sum_{k=0}^{\ell_0-1}
\prod_{m=1}^{K^{\rm{II}}}\frac{\mathscr{X}^{k,0}_k(q,\lambda_m)}{\mathscr{Y}^{k,0;1}_{k;1}(q,\lambda_m)}\prod_{i=1}^{K^{\rm{I}}}\mathscr{Y}^{k,0;1}_{k;1}(q,p_i)\Lambda^{(6v)}(u_0+{\textstyle{\frac{\ell_0-1}{2}}}-k,\vec{u}_\lambda
),
\end{eqnarray}
where again $\Lambda^{(6v)}$ is the eigenvalue of the auxiliary
6-vertex model, and $\vec{u}_\lambda  = (u_{\lambda_1 }, \cdots,
u_{\lambda_{K^{\rm II}}} )$. The auxiliary roots satisfy the
following equations
\begin{eqnarray}
\Lambda^{(6v)}(u_{\lambda_j } ,\vec{u}_\lambda )\prod_{i=1}^{K^{\rm{I}}}\mathscr{Y}^{0,0;1}_{0;1}(\lambda_j,p_i)&=&1,\\
\prod_{i=1}^{K^{\rm{II}}}b(w_j,u_{\lambda_i} )\prod_{i=1,i\neq
j}^{K^{\rm{III}}} \frac{b(w_i,w_j)}{b(w_j,w_i)}&=&1.
\end{eqnarray}
In appendix \ref{sec;TasBA} we give a full derivation of this
eigenvalue and auxiliary equations for the case $K^{\rm{III}}=0$.
We would also like to mention that the form of the
eigenvalues appears in the form of factorized
products of single-excitation terms - a somewhat expected
feature, which makes us more confident about the
generalization procedure.

We would like to point out that the dependence of the auxiliary
parameters $\lambda_m$ only appears in the form $x^+(\lambda_m)$.
In order to compare with the known Bethe equations we relabel this
to be $x^+(\lambda_m) \equiv y_m$. We also rescale
$w_i\longrightarrow \frac{2i}{g}w_i$.
In terms of these parameters, the eigenvalues become
\begin{eqnarray}\label{eqn;FullEignvalue}
&&\Lambda(q|\vec{p})=\prod_{i=1}^{K^{\rm{II}}}{\textstyle{\frac{y_i-x^-_0}{y_i-x^+_0}\sqrt{\frac{x^+_0}{x^-_0}}
+}}\\
&&
{\textstyle{+}}\prod_{i=1}^{K^{\rm{II}}}{\textstyle{\frac{y_i-x^-_0}{y_i-x^+_0}\sqrt{\frac{x^+_0}{x^-_0}}
\left[\frac{x^+_0+\frac{1}{x^+_0}-y_i-\frac{1}{y_i}}{x^+_0+\frac{1}{x^+_0}-y_i-\frac{1}{y_i}-\frac{2i\ell_0}{g}}\right]}}
\prod_{i=1}^{K^{\rm{I}}}
{\textstyle{\left[\frac{(x^-_0-x^-_i)(1-x^-_0 x^+_i)}{(x^-_0-x^+_i)(1-x^+_0 x^+_i)}
\sqrt{\frac{x^+_0x^+_i}{x^-_0x^-_i}}\mathscr{X}^{\ell_0,0}_{\ell_0}\right]}}
\nonumber\\
&&{\textstyle{+}}
\sum_{k=1}^{\ell_0-1}\prod_{i=1}^{K^{\rm{II}}}{\textstyle{\frac{y_i-x^-_0}{y_i-x^+_0}\sqrt{\frac{x^+_0}{x^-_0}}
\left[\frac{x^+_0+\frac{1}{x^+_0}-y_i-\frac{1}{y_i}}{x^+_0+\frac{1}{x^+_0}-y_i-\frac{1}{y_i}-\frac{2ik}{g}}\right]}}
\left\{\prod_{i=1}^{K^{\rm{I}}}{\textstyle{\lambda_+(q,p_i,k)+}}\right.\left.\prod_{i=1}^{K^{\rm{I}}}{\textstyle{\lambda_-(q,p_i,k)}}\right\}\nonumber\\
&&\quad
{\textstyle{-}}\sum_{k=0}^{\ell_0-1}\prod_{i=1}^{K^{\rm{II}}}{\textstyle{\frac{y_i-x^-_0}{y_i-x^+_0}\sqrt{\frac{x^+_0}{x^-_0}}
\left[\frac{x^+_0+\frac{1}{x^+_0}-y_i-\frac{1}{y_i}}{x^+_0+\frac{1}{x^+_0}-y_i-\frac{1}{y_i}-\frac{2ik}{g}}\right]}}
\prod_{i=1}^{K^{\rm{I}}}{\textstyle{\frac{x^+_0-x^+_i}{x^-_0-x^+_i}\sqrt{\frac{x^-_0}{x^+_0}}\left[1-\frac{k}{u_0-u_i+\frac{\ell_0-\ell_i}{2}
}\right]}}\times\nonumber\\
&&\quad\times
{\textstyle{\mathscr{X}^{k,0}_k}}\left\{\prod_{i=1}^{K^{\rm{III}}}{\textstyle{\frac{w_i-x^+_0-\frac{1}{x^+_0}+\frac{i(2k-1)}{g}}{w_i-x^+_0-\frac{1}{x^+_0}+\frac{i(2k+1)}{g}}+
}}\prod_{i=1}^{K^{\rm{II}}}{\textstyle{\frac{y_i+\frac{1}{y_i}-x^+_0-\frac{1}{x^+_0}+\frac{2ik}{g}}{y_i+\frac{1}{y_i}-x^+_0-\frac{1}{x^+_0}+\frac{2i(k+1)}{g}}}}\prod_{i=1}^{K^{\rm{III}}}{\textstyle{\frac{w_i-x^+_0-\frac{1}{x^+_0}+\frac{i(2k+3)}{g}}{w_i-x^+_0-\frac{1}{x^+_0}+\frac{i(2k+1)}{g}}}}\right\}.\nonumber
\end{eqnarray}
and the above auxiliary Bethe equations transform into the well-known ones:
\begin{eqnarray}
\label{bennote}
\prod_{i=1}^{K^{\rm{I}}}\frac{y_k-x^-_i}{y_k-x^+_i}\sqrt{\frac{x^+_i}{x^-_i}}&=&
\prod_{i=1}^{K^{\rm{III}}}\frac{w_i-y_k-\frac{1}{y_k}-\frac{i}{g}}{w_i-y_k-\frac{1}{y_k}+\frac{i}{g}},\\
\prod_{i=1}^{K^{\rm{II}}}\frac{w_k-y_i-\frac{1}{y_i}+\frac{i}{g}}{w_k-y_i-\frac{1}{y_i}-\frac{i}{g}}
&=& \prod_{i=1,i\neq
k}^{K^{\rm{III}}}\frac{w_k-w_i+\frac{2i}{g}}{w_k-w_i-\frac{2i}{g}}.\nonumber
\end{eqnarray}
Once again, we find that for {\it all} fundamental representations (including the auxiliary space) this
agrees with what obtained in \cite{Martins:2007hb}. Analogously to formula (41) from the
same paper, one can derive the complete set of Bethe equations
from the transfer matrix. One finds that the one-particle momenta
should satisfy
\begin{eqnarray}
e^{ip_j L} =  \Lambda(p_j|\vec{p}).
\end{eqnarray}
The first thing one should notice is that if $q=p_j$ and
$\ell_0=\ell_j$, then $\mathscr{X}^{k,0}_k=0$ if $k>0$. This means
that the only surviving terms is found to be the first one. This
gives the following Bethe equations (after explicitly including the appropriate
scalar factor $S_0$)
\begin{eqnarray}
e^{ip_j L} =  \prod_{i=1,i\neq j}^{K^{\rm{I}}} S_0(p_j,p_i)
\prod_{m=1}^{K^{\rm{II}}}
\frac{y_m-x^-_j}{y_m-x^+_j}\sqrt{\frac{x^+_j}{x^-_j}}.
\end{eqnarray}
Together with the above set of auxiliary Bethe equations, these
indeed agree with the fused Bethe equations.

\section{Different vacua and fusion}

In the previous sections we deduced the spectrum of the transfer
matrix. We found all of its eigenvalues, characterized by the
numbers $K^{\rm{I,II,III}}$. The eigenvalues were obtained by starting with a
vacuum with numbers $K^{\rm{II}}=K^{\rm{III}}=0$, which proved to
be an eigenstate, and then applying creation operators that generate eigenstates with different quantum
numbers. Of course, our choice of vacuum is not unique. We can
build up our algebraic Bethe ansatz starting from a different
vacuum. One trivial example of this would be to start with $w_2$
instead of $w_1$. A more interesting case arises when all
physical particles are fermions.

\subsection{$\alg{sl}(2)$ vacuum}

Consider a fermionic vacuum with all the physical particles in the fundamental representation:
\begin{eqnarray}
|0\rangle_P^{\prime}= \theta_{3}\otimes\ldots\otimes\theta_{3}.
\end{eqnarray}
This vacuum has quantum numbers $K^{\rm{II}}=K^{\rm{I}}$ and
$K^{\rm{III}}=0$. One can easily check that this vacuum is
also an eigenstate. The action of the diagonal elements of fermionic type
of the transfer matrix is given by:
\begin{eqnarray}
\begin{aligned}
\mathcal{T}_{3;k}^{3;k}|0\rangle_P^{\prime} &=& \prod_{i=1}^{K^{\rm{I}}}\mathscr{X}^{k,0}_k(q,p_i)|0\rangle_P^{\prime},\nonumber\\
\mathcal{T}_{4;k}^{4;k}|0\rangle_P^{\prime} &=&
\prod_{i=1}^{K^{\rm{I}}}\mathscr{Z}^{k,0;6}_{k;6}(q,p_i)|0\rangle_P^{\prime}.
\end{aligned}
\end{eqnarray}
The explicit values for these scattering elements is given in
appendix \ref{A}, and one obtains
\begin{eqnarray}
\label{frmi}
\begin{aligned}
\mathcal{T}_{3;k}^{3;k}|0\rangle_P^{\prime} &=
\prod_{i=1}^{K^{\rm{I}}}\frac{x_0^--x_i^+}{x^+_0-x^-_i}\sqrt{\frac{x^+_0x^-_i}{x^-_0x^+_i}}|0\rangle_P^{\prime},
\\
\mathcal{T}_{4;k}^{4;k}|0\rangle_P^{\prime} &=
\prod_{i=1}^{K^{\rm{I}}}\frac{x_0^--x_i^-}{x^+_0-x^-_i}\frac{x_i^--
\frac{1}{x_0^+}}{x_i^+-\frac{1}{x_0^+}}\sqrt{\frac{x^+_0x^+_i}{x^-_0x^-_i}}|0\rangle_P^{\prime}.
\end{aligned}
\end{eqnarray}
Notice that these elements are \emph{independent} of $k$. This
means that, when summing over $k$, this will only give a factor of
$\ell_0$.

The next step is to consider the bosonic elements
$\mathcal{T}^k_k,\mathcal{T}^{34,k}_{34,k}$. Let us
again split off the contributions from $k=0$ and $k=\ell_0$. The corresponding elements
$\mathcal{T}^0_0,\mathcal{T}^{\ell_0}_{\ell_0}$ act on this new vacuum as
\begin{eqnarray}
\mathcal{T}^0_0|0\rangle_P^{\prime}=\mathcal{T}^{\ell_0}_{\ell_0}|0\rangle_P^{\prime}&=&\prod_{i=1}^{K^{\rm{I}}}\mathscr{Y}^{k,0;2}_{k;2}|0\rangle_P^{\prime},\nonumber\\
&=&\prod_{i=1}^{K^{\rm{I}}}\frac{x_0^--x^-_i}{x_0^+-x^-_i}\sqrt{\frac{x^+_0}{x^-_0}}|0\rangle_P^{\prime}.
\end{eqnarray}
For the remaining elements one finds again, as in the case of the vacuum we have been using before, an additional matrix
structure. More precisely, this time one needs to compute the
eigenvalues of the matrix
\begin{eqnarray}
\begin{pmatrix}
\mathscr{Y}^{k,0;2}_{k;2} & \mathscr{Y}^{k,0;4}_{k;2}\\
\mathscr{Y}^{k,0;2}_{k;4} & \mathscr{Y}^{k,0;4}_{k;4}
\end{pmatrix}.
\end{eqnarray}
However, here one encounters the remarkable fact that
$\mathscr{Y}^{k,0;4}_{k;2}=\mathscr{Y}^{k,0;2}_{k;4}=0$, and the
matrix is therefore already diagonal. Hence, the eigenvalues are easily read
off, and one finds
\begin{eqnarray}
\mathcal{T}^k_k|0\rangle_P^{\prime}&=&\prod_{i=1}^{K^{\rm{I}}}\mathscr{Y}^{k,0;2}_{k;2}(q,p_i) |0\rangle_P^{\prime}
=\prod_{i=1}^{K^{\rm{I}}}\frac{x_0^--x^-_i}{x_0^+-x^-_i}\sqrt{\frac{x^+_0}{x^-_0}}|0\rangle_P^{\prime}
\end{eqnarray}
and
\begin{eqnarray}
\mathcal{T}^{\alpha4,k}_{\alpha4,k}|0\rangle_P^{\prime}&=&\prod_{i=1}^{K^{\rm{I}}}
\mathscr{Y}^{k,0;4}_{k;4}(q,p_i) |0\rangle_P^{\prime}
=\prod_{i=1}^{K^{\rm{I}}}\frac{x_0^--x^+_i}{x_0^+-x^-_i}\frac{x_i^--\frac{1}{x_0^+}}{x_i^+-\frac{1}{x_0^+}}\sqrt{\frac{x^+_0}{x^-_0}}|0\rangle_P^{\prime}.
\end{eqnarray}
Similarly to the fermionic contributions (\ref{frmi}), and once
again in contrast to the bosonic vacuum, one finds that these
terms are \emph{independent} of $k$. Summing everything up finally
gives that $|0\rangle_P^{\prime}$ is an eigenvalue of the transfer
matrix with eigenvalue
\begin{eqnarray}
\label{anti}
\Lambda(q|\vec{p})=&&(\ell_0+1)\prod_{i=1}^{K^{\rm{I}}}\frac{x_0^--x^-_i}{x_0^+-x^-_i}
\sqrt{\frac{x^+_0}{x^-_0}}-
\ell_0\prod_{i=1}^{K^{\rm{I}}}\frac{x_0^--x_i^+}{x^+_0-x^-_i}
\sqrt{\frac{x^+_0x^-_i}{x^-_0x^+_i}}\, -\\\nonumber &&-\ell_0
\prod_{i=1}^{K^{\rm{I}}}\frac{x_0^--x_i^-}{x^+_0-x^-_i}\frac{x_i^--\frac{1}{x_0^+}}
{x_i^+-\frac{1}{x_0^+}}\sqrt{\frac{x^+_0x^+_i}{x^-_0x^-_i}}
+(\ell_0-1)\prod_{i=1}^{K^{\rm{I}}}\frac{x_0^--x^+_i}{x_0^+-x^-_i}
\frac{x_i^--\frac{1}{x_0^+}}{x_i^+-\frac{1}{x_0^+}}\sqrt{\frac{x^+_0}{x^-_0}}.
\end{eqnarray}
This precisely agrees with the result of \cite{Beisert:2006qh} for antisymmetric representations.

Let us remark that the spectrum is clearly independent of the
choice of vacuum. Hence, one should find the same eigenvalues when
starting from this or from the bosonic vacuum, that we used in
this paper, provided one excites the appropriate set of auxiliary
roots. In particular, in the bosonic vacuum one has first to solve
the $K^{\rm{II}}$ auxiliary BAE, and then use these solutions to
find the corresponding eigenvalue, which should therefore agree
with (\ref{anti}). In fact, conversion of one eigenvalue into the
other can be obtained by means of duality transformations
\cite{Beisert:2005di}. We would also like to notice that the
result obtained in this section for fundamental representations in
the physical space happens to have nice fusion properties, and one
can think of combining several of such elementary transfer
matrices to obtain more general ones. This approach has been
followed for instance in \cite{Hatsuda:2008gd}.

\subsection{$\alg{su}(2)$ vacuum}

Let us now come back to the bosonic vacuum we have been using
throughout the paper in the derivation of the ABA. In
\cite{Beisert:2006qh}, a prescription for computing the transfer
matrix eigenvalues on the $\alg{su}(2)$ vacuum (symmetric
representation), for all physical legs in the fundamental
representation, was also given. The formula was expressed in terms
of an expansion of the inverse of a quantum characteristic
function. We have found that this prescription indeed produces the
same eigenvalues as obtained from our general formula
(\ref{eqn;FullEignvalue}), when restricting the latter to
fundamental particles in the physical space. To this purpose, we
explicitly work out here below the above mentioned expansion
following \cite{Beisert:2006qh}, adapting the calculation to the
notations we use in this paper. We will then compare the final
formula with the suitable restriction of our result
(\ref{eqn;FullEignvalue}), finding perfect agreement. Indeed, we
will be able to relax the condition of physical legs in the
fundamental representation, by making the conjectured expression
for the quantum characteristic function slightly more general. We
will then find agreement with such a formula in the general case
$\ell_i \neq 1$ as well. \ad{All these agreements} will however be reached in
a quite non-trivial and interesting fashion. We therefore think
that the explicit calculation we reproduce in what follows will
precisely help clarifying this fact.

\bigskip
Following \cite{Beisert:2006qh}, we define the shift operator $U$ by
\begin{equation}
U \ssp f (u) \, U^{-1} = f \( u + \frac{1}{2} \),
\label{def:shift ops}
\end{equation}
and introduce the notation
\begin{equation}
f^{[\ell]} (u) \equiv U^\ell \ssp f (u) \, U^{-\ell} = f \( u + \frac{\ell}{2} \).
\end{equation}
The spectral parameters of an elementary particle, defined in \eqref{uvar}, satisfy the relation
\begin{equation}
x^{[1]} + \frac{1}{x^{[1]}} - x^{[-1]} - \frac{1}{x^{[-1]}} = \frac{2i}{g} \,.
\label{def:torus}
\end{equation}
By successive applications of the shift operator to \eqref{def:torus}, one finds that the pair of variables $\{ x^{[\ell]}, x^{[\ell-2k]} \}$ defines another rapidity torus
\begin{equation}
x^{[\ell]} + \frac{1}{x^{[\ell]}} - x^{[\ell-2k]} - \frac{1}{x^{[\ell-2k]}} = \frac{2ik}{g} \,.
\end{equation}
There are two choices of branch for $x_a^{[\ell-2k]}$ for a given $x^{[\ell]}$, as can be seen by
\begin{equation}
x^{[\ell-2k]} = \frac12 \( x^{[\ell]} + \frac{1}{x^{[\ell]}} - \frac{2 i k}{g} + \sqrt{ \( x^{[\ell]} + \frac{1}{x^{[\ell]}} - \frac{2 i k}{g} \)^2 - 4} \; \).
\label{x intermediate branches}
\end{equation}
We also use $y_i + 1/y_i \ad{= i v_i}$ in what follows.\footnote{\ad{Interestingly, the final result \eqref{symmetric transfer 4} is almost invariant under the map $y_i \mapsto 1/y_i$, except for an overall factor.}}

Let $\vev{\ell_0-1,0}$ be the $\ell_0$\,-th symmetric representation of $\alg{su}(2|2)$. The conjecture states that the transfer matrix for such a representation $T_{\vev{\ell_0-1,0}} (u_0|\{\vec u, \vec v, \vec w\})$ is generated by $T_{\vev{0,0}} (u_0|\{\vec u, \vec v, \vec w\})$, where the generating function is equal to the inverse of the quantum characteristic function:
\begin{alignat}{3}
D_0^{-1} &\defeq \( 1 - U_0 T_4 U_0 \)^{-1} \( 1 - U_0 T_3 U_0 \) \( 1 - U_0 T_2 U_0 \) \( 1 - U_0 T_1 U_0 \)^{-1} , \\[1mm]
&= \( 1 + \sum_{h=1}^\infty (U_0 T_4 U_0)^h \) \( 1 - U_0 T_3 U_0 \) \( 1 - U_0 T_2 U_0 \) \( 1 + \sum_{k=1}^\infty (U_0 T_1 U_0)^k \),
\notag \\[1mm]
&\equiv \sum_{\ell_0=0}^\infty U_0^{\ell_0} \, T_{\vev{\ell_0-1,0}} (u_0|\{\vec u, \vec v, \vec w\}) \, U_0^{\ell_0} \,.
\label{symmetric transfer 1}
\end{alignat}
Here $U_0$ is the shift operator for $u_0$\,. The first few terms can be found as follows:
\begin{alignat}{3}
D_0^{-1} &= 1 + U_0 \( T_4 - T_3 - T_2 + T_1 \) U_0
\label{qch expansion 1} \\
&\hspace{-5mm} + U_0^2 \Bigl\{ T_4^{[-1]} T_4^{[1]} + T_4^{[-1]} T_1^{[1]} + T_1^{[-1]} T_1^{[1]} + T_3^{[-1]} T_2^{[1]} \notag \\
&\hspace{40mm} - T_4^{[-1]} (T_3^{[1]} + T_2^{[1]}) - (T_3^{[-1]} + T_2^{[-1]}) T_1^{[1]} \Bigr\} U_0^2 \notag \\[1mm]
&\hspace{-5mm} + U_0^3 \, \Bigl\{ T_4^{[-2]} T_4 T_4^{[+2]} + T_4^{[-2]} T_4 T_1^{[+2]} + T_4^{[-2]} T_1 T_1^{[+2]} + T_1^{[-2]} T_1 T_1^{[+2]} + T_4^{[-2]} T_3 T_2^{[+2]}\notag \\[1mm]
&\hspace{-5mm} + T_3^{[-2]} T_2 T_1^{[+2]} - T_4^{[-2]} T_4 \( T_3^{[+2]} + T_2^{[+2]} \) - T_4^{[-2]} \( T_3 + T_2 \) T_1^{[+2]} \notag \\
&\hspace{-5mm} - \( T_3^{[-2]} + T_2^{[-2]} \) T_1 T_1^{[+2]} \Bigr\} \, U_0^3
\ + \ \cdots,
\end{alignat}
and, in general,
\begin{equation}
T_{\vev{\ell_0-1,0}} (u_0|\{\vec u, \vec v, \vec w\}) = \tau_{\ell_0,0} - \tau_{\ell_0,1} \[ T_3 \] - \tau_{\ell_0,1} \[ T_2 \] + \tau_{\ell_0,2} \[ T_3 \,, T_2 \],
\label{symmetric transfer 2}
\end{equation}
where
\begin{alignat}{5}
\tau_{\ell_0,0} &= \sum_{k=0}^{\ell_0} T_4^{[-\ell_0+1]} T_4^{[-\ell_0+3]} \cdots T_4^{[\ell_0-2k-3]} \, T_1^{[\ell_0-2k-1]} \cdots T_1^{[\ell_0-1]} \,,
\label{tau m0} \\[1mm]
\tau_{\ell_0,1} \left[ X \right] &= \sum_{k=0}^{\ell_0-1} T_4^{[-\ell_0+1]} T_4^{[-\ell_0+3]} \cdots T_4^{[-\ell_0+2k-1]} \ \times
\notag \\[1mm]
&\hspace{50mm} X^{[\ell_0-2k-1]} \, T_1^{[\ell_0-2k+1]} \cdots T_1^{[\ell_0-1]} \,,
\label{tau m1} \\[1mm]
\tau_{\ell_0,2} \left[ X, Y \right] &= \sum_{k=0}^{\ell_0-2} T_4^{[-\ell_0+1]} T_4^{[-\ell_0+3]} \cdots T_4^{[-\ell_0+2k-1]} \ \times
\notag \\[1mm]
&\hspace{35mm} X^{[\ell_0-2k-3]} \, Y^{[\ell_0-2k-1]} \, T_1^{[\ell_0-2k+1]} \cdots T_1^{[\ell_0-1]} \,.
\label{tau m2}
\end{alignat}
The first line of \eqref{qch expansion 1} gives the transfer matrix for the fundamental representation as
\begin{equation}
T_{\vev{0,0}} (u_0|\{\vec u, \vec v, \vec w\}) = T_1 - T_2 - T_3 + T_4.
\label{T00 expand}
\end{equation}
We recall that the left hand side of this equation is given explicitly by \eqref{eqn;FullEignvalue} at $\ell_0=1$, which reads
\begin{align}
\Lambda(q|\vec{p}) &= \prod_{i=1}^{K^{\rm{II}}}{\textstyle{\frac{y_i-x^-_0}{y_i-x^+_0}\sqrt{\frac{x^+_0}{x^-_0}}
+}}\\
&{\textstyle{+}}
\prod_{i=1}^{K^{\rm{II}}}{\textstyle{\frac{y_i-x^-_0}{y_i-x^+_0}\sqrt{\frac{x^+_0}{x^-_0}}\left[
\frac{x^+_0+\frac{1}{x^+_0}-y_i-\frac{1}{y_i}}{x^+_0+\frac{1}{x^+_0}-y_i-\frac{1}{y_i}-\frac{2i}{g}} \right]}} \prod_{i=1}^{K^{\rm{I}}}
{\textstyle{\left[\frac{(x^-_0-x^-_i)(1-x^-_0 x^+_i)}{(x^-_0-x^+_i)(1-x^+_0 x^+_i)}
\sqrt{\frac{x^+_0 x^+_i}{x^-_0 x^-_i}}\mathscr{X}^{1,0}_{1}\right]}}
\nonumber\\
&\qquad
- \prod_{i=1}^{K^{\rm{II}}}{\textstyle{\frac{y_i-x^-_0}{y_i-x^+_0}\sqrt{\frac{x^+_0}{x^-_0}} }}\prod_{i=1}^{K^{\rm{I}}}{\textstyle{\frac{x^+_0-x^+_i}{x^-_0-x^+_i}\sqrt{\frac{x^-_0}{x^+_0}} }}\times\nonumber\\
&\qquad \times
{\textstyle{\mathscr{X}^{0,0}_0}}\left\{\prod_{i=1}^{K^{\rm{III}}}{\textstyle{\frac{w_i-x^+_0-\frac{1}{x^+_0}-\frac{i}{g}}{w_i-x^+_0-\frac{1}{x^+_0}+\frac{i}{g}}+
}}\prod_{i=1}^{K^{\rm{II}}}{\textstyle{\frac{y_i+\frac{1}{y_i}-x^+_0-\frac{1}{x^+_0}}{y_i+\frac{1}{y_i}-x^+_0-\frac{1}{x^+_0}+\frac{2i}{g}}}}\prod_{i=1}^{K^{\rm{III}}}{\textstyle{\frac{w_i-x^+_0-\frac{1}{x^+_0}+\frac{3i}{g}}{w_i-x^+_0-\frac{1}{x^+_0}+\frac{i}{g}}}}\right\}.\nonumber
\end{align}
Therefore, $\Lambda(q|\vec{p})$ may be equated with the right hand side of \eqref{T00 expand} term by term.
We simplify the above expression of $\Lambda(q|\vec{p})$ by introducing variables ${\sf w}_i$ and ${\sf v}_i$ as follows\footnote{Our notation is $x_0^\pm = x^{[\pm \ell_0]}$, and $\ell_0=1$ is used when discussing the fundamental transfer matrix. Note that the shift operator does not act on $x_i^\pm$.}:
\begin{alignat}{3}
w_i - x_0^\pm - \frac{1}{x_0^\pm} + \frac{in}{g} &\equiv \( {\sf w}_i - u_0 + \frac{\mp \ell_0 + n}{2} \) \frac{2i}{g} \,,
\\[1mm]
y_i + \frac{1}{y_i} - x_0^\pm - \frac{1}{x_0^\pm} + \frac{in}{g} &\equiv \( {\sf v}_i - u_0 + \frac{\mp \ell_0 + n}{2} \) \frac{2i}{g} \,.
\end{alignat}
\ad{With the help of \eqref{xlolook} and}
\begin{equation}
\mathscr{X}^{1,0}_1 = \frac{u_0 - u_i + \frac{1 - \ell_i}{2}}{u_0
- u_i + \frac{\ell_i - 1}{2}} \, \mathcal{D} = \frac{\( x_0^+ -
x_i^+ \) \(1 - \frac{1}{x_0^+ x_i^+} \) \( x_0^- - x_i^+ \)}{\(
x_0^- - x_i^- \) \(1 - \frac{1}{x_0^- x_i^-} \) \( x_0^+ - x_i^-
\)} \sqrt{\frac{x_0^+ x_i^-}{x_0^- x_i^+} } \,,
\end{equation}
\ad{it} produces
\begin{eqnarray}
&&\Lambda(q|\vec{p}) = \prod_{i=1}^{K^{\rm{II}}} \frac{y_i-x^-_0}{y_i-x^+_0}\sqrt{\frac{x^+_0}{x^-_0}}
\label{Full fundamental}
\\[1mm]
&&
+ \prod_{i=1}^{K^{\rm{II}}} \frac{y_i-x^-_0}{y_i-x^+_0}\sqrt{\frac{x^+_0}{x^-_0}} \left[
\frac{{\sf v}_i - u_0 - \frac12}{{\sf v}_i - u_0 + \frac12} \right]
\prod_{i=1}^{K^{\rm{I}}}
\frac{(x^+_0 - x^+_i) \(1 - \frac{1}{x_0^- x_i^+}\)}{(x^+_0 - x^-_i) \(1 - \frac{1}{x_0^- x_i^-}\)}
\nonumber\\[1mm]
&&\quad
- \prod_{i=1}^{K^{\rm{II}}} \frac{y_i-x^-_0}{y_i-x^+_0}\sqrt{\frac{x^+_0}{x^-_0}}
\prod_{i=1}^{K^{\rm{I}}} \frac{x^+_0-x^+_i}{x_0^+ - x_i^-} \sqrt{\frac{x_i^-}{x_i^+}} \times
\nonumber\\[1mm]
&&\hspace{50mm} \times
\left\{\prod_{i=1}^{K^{\rm{III}}} \frac{{\sf w}_i - u_0 -1}{{\sf w}_i - u_0} +
\prod_{i=1}^{K^{\rm{II}}} \frac{{\sf v}_i - u_0 - \frac12}{{\sf v}_i - u_0 + \frac12}
\prod_{i=1}^{K^{\rm{III}}} \frac{{\sf w}_i - u_0 + 1}{{\sf w}_i - u_0} \right\}.
\nonumber
\end{eqnarray}
It is useful to separate a common factor in the following fashion:
\begin{equation}
T_i = S_{\vev{0,0}} \, \tilde T_i \,,\qquad
S_{\vev{0,0}} \equiv \prod_{i=1}^{K^{\rm{II}}} \frac{y_i-x^-_0}{y_i-x^+_0}\sqrt{\frac{x^+_0}{x^-_0}} \,,\qquad (i=1, \ldots , 4).
\label{def:tilde Ti}
\end{equation}
Then, the tilded functions can be written as
\begin{alignat}{3}
\tilde T_1 &= \prod_{i=1}^{K^{\rm{II}}}
\frac{{\sf v}_i - u_0 - \frac12}{{\sf v}_i - u_0 + \frac12} \,
\prod_{i=1}^{K^{\rm{I}}}
\frac{\(1 - \frac{1}{x_0^- x_i^+}\) (x^+_0 - x^+_i)}{\(1 - \frac{1}{x_0^- x_i^-}\) (x^+_0 - x^-_i)} \,,
\label{T1 Full} \\[1mm]
\tilde T_2 &= \prod_{i=1}^{K^{\rm{III}}} \frac{{\sf w}_i - u_0 + 1}{{\sf w}_i - u_0} \,
\prod_{i=1}^{K^{\rm{II}}} \frac{{\sf v}_i - u_0 - \frac12}{{\sf v}_i - u_0 + \frac12} \,
\prod_{i=1}^{K^{\rm{I}}} \frac{x^+_0-x^+_i}{x_0^+ - x_i^-} \sqrt{\frac{x_i^-}{x_i^+}} \,,
\label{T2 Full} \\[2mm]
\tilde T_3 &= \prod_{i=1}^{K^{\rm{III}}} \frac{{\sf w}_i - u_0 -1}{{\sf w}_i - u_0} \,
\prod_{i=1}^{K^{\rm{I}}} \frac{x^+_0-x^+_i}{x_0^+ - x_i^-} \sqrt{\frac{x_i^-}{x_i^+}} \,,
\label{T3 Full} \\[1mm]
\tilde T_4 &= 1.
\label{T4 Full}
\end{alignat}
Note that different identification of $\tilde T_i$'s would produce the transfer matrix for different representations \cite{Beisert:2005di}.

Let us evaluate the function $\tau$'s appearing in the conjectured transfer matrix for the $\ell_0$\,-th symmetric representation \eqref{symmetric transfer 2}. They can be simplified by using
\begin{alignat}{3}
\tilde T_1^{[\ell_0-1]} &= \prod_{i=1}^{K^{\rm{II}}}
\frac{{\sf v}_i - u_0 - \frac{\ell_0}{2}}{{\sf v}_i - u_0 - \frac{\ell_0-2}{2}} \,
\prod_{i=1}^{K^{\rm{I}}} \[ \frac{1-\frac{1}{x^{[\ell_0-2]}_0 x^+_i}}{1-\frac{1}{x^{[\ell_0-2]}_0 x^-_i}} \, \frac{x^{[\ell_0]}_0-x^+_i}{x^{[\ell_0]}_0-x^-_i} \],
\notag \\[1mm]
&= \prod_{i=1}^{K^{\rm{II}}}
\frac{{\sf v}_i - u_0 - \frac{\ell_0}{2}}{{\sf v}_i - u_0 - \frac{\ell_0-2}{2}} \,
\prod_{i=1}^{K^{\rm{I}}} \[
\frac{x^{[\ell_0-2]}_0 - x^-_i}{x^{[\ell_0-2]}_0 - x^+_i} \,
\frac{u_0 - u_i + \frac{\ell_0 - \ell_i - 2}{2}}{u_0 - u_i + \frac{\ell_0 + \ell_i - 2}{2}} \,
\frac{x^{[\ell_0]}_0-x^+_i}{x^{[\ell_0]}_0-x^-_i} \],
\notag \\[1mm]
&= \prod_{i=1}^{K^{\rm{II}}}
\frac{{\sf v}_i - u_0 - \frac{\ell_0}{2}}{{\sf v}_i - u_0 - \frac{\ell_0-2}{2}} \ \times
\notag \\[1mm]
&\qquad \prod_{i=1}^{K^{\rm{I}}} \[ \frac{x^{[\ell_0-2]}_0 - x^-_i}{x^{[\ell_0-2]}_0 - x^+_i} \, \frac{u_0 - u_i + \frac{\ell_0 - \ell_i - 2}{2}}{u_0 - u_i + \frac{\ell_0 + \ell_i - 2}{2}} \,
\frac{u_0 - u_i + \frac{\ell_0 - \ell_i}{2}}{u_0 - u_i + \frac{\ell_0 + \ell_i}{2}} \,
\frac{1 - \frac{1}{x^{[\ell_0]}_0 x^-_i}}{1 - \frac{1}{x^{[\ell_0]}_0 x^+_i}} \]
\end{alignat}
and, therefore,
\begin{multline}
\tilde T_1^{[\ell_0-3]} \, \tilde T_1^{[\ell_0-1]} = \prod_{i=1}^{K^{\rm{II}}}
\frac{{\sf v}_i - u_0 - \frac{\ell_0}{2}}{{\sf v}_i - u_0 - \frac{\ell_0-4}{2}} \ \times
\\[1mm]
\prod_{i=1}^{K^{\rm{I}}} \[
\frac{x^{[\ell_0-4]}_0 - x^-_i}{x^{[\ell_0-4]}_0 - x^+_i} \,
\frac{u_0 - u_i + \frac{\ell_0 - \ell_i - 4}{2}}{u_0 - u_i + \frac{\ell_0 + \ell_i - 4}{2}} \,
\frac{u_0 - u_i + \frac{\ell_0 - \ell_i - 2}{2}}{u_0 - u_i + \frac{\ell_0 + \ell_i - 2}{2}} \,
\frac{x^{[\ell_0]}_0-x^+_i}{x^{[\ell_0]}_0-x^-_i} \].
\end{multline}
Thus, $\tau_{\ell_0,0}$ becomes
\begin{align}
&\tau_{\ell_0,0} = S_{\vev{\ell_0-1,0}} \ \times
\label{symm tau0} \\[1mm]
&\quad \( 1 + \sum_{k=1}^{\ell_0} \pare{
\prod_{i=1}^{K^{\rm{II}}}
\frac{{\sf v}_i - u_0 - \frac{\ell_0}{2}}{{\sf v}_i - u_0 - \frac{\ell_0-2k}{2}} \,
\prod_{i=1}^{K^{\rm{I}}} \left[
\frac{x^{[\ell_0-2k]}_0 - x^-_i}{x^{[\ell_0-2k]}_0 - x^+_i} \,
\, X_1^k \(u_0 - u_i \,, \ell_0 \,, \ell_i\)
\frac{x^{[\ell_0]}_0-x^+_i}{x^{[\ell_0]}_0-x^-_i} \right] } \),
\notag
\end{align}
where we have introduced
\begin{align}
&S_{\vev{\ell_0-1,0}} = \prod_{k=1}^{\ell_0} S_{\vev{0,0}}^{[\ell_0+1-2k]}
= \prod_{i=1}^{K^{\rm{II}}} \frac{y_i-x^{[-\ell_0]}_0}{y_i - x^{[\ell_0]}_0}\sqrt{\frac{x^{[\ell_0]}_0}{x^{[-\ell_0]}_0}} \,,
\label{def:S ell0} \\
&X_l^k \(u, \ell_0 \,, \ell_i\) = \prod_{j=l}^k \frac{u + \frac{\ell_0 - \ell_i - 2j}{2}}{u + \frac{\ell_0 + \ell_i - 2j}{2}} \,.
\label{def:Xik}
\end{align}
Similarly, $\tau_{\ell_0,1} \[T_3\], \tau_{\ell_0,1} \[T_2\]$ and
$\ad{\tau_{\ell_0,2} \[T_3 \,, T_2 \]}$ are given by
\begin{alignat}{3}
\tau_{\ell_0,1} \[ T_3 \] &= S_{\vev{\ell_0-1,0}} \, \sum_{k=0}^{\ell_0-1} \Biggl\{
\prod_{i=1}^{K^{\rm{III}}} \frac{{\sf w}_i - u_0 - \frac{\ell_0-2k+1}{2}}{{\sf w}_i - u_0 - \frac{\ell_0-2k-1}{2}} \,
\prod_{i=1}^{K^{\rm{II}}} \frac{{\sf v}_i - u_0 - \frac{\ell_0}{2}}{{\sf v}_i - u_0 - \frac{\ell_0-2k}{2}} \ \times
\label{symm tau1 T3} \\[1mm]
&\hspace{50mm}
\prod_{i=1}^{K^{\rm{I}}} \[
\sqrt{\frac{x^-_i}{x^+_i}} \,
X_1^k \(u_0 - u_i \,, \ell_0 \,, \ell_i\)
\frac{x^{[\ell_0]}_0-x^+_i}{x^{[\ell_0]}_0-x^-_i} \] \Biggr\},
\notag \\[2mm]
\tau_{\ell_0,1} \left[ T_2 \right] &= S_{\vev{\ell_0-1,0}} \, \sum_{k=0}^{\ell_0-1} \Biggl\{
\prod_{i=1}^{K^{\rm{III}}} \frac{{\sf w}_i - u_0 - \frac{\ell_0-2k-3}{2}}{{\sf w}_i - u_0 - \frac{\ell_0-2k-1}{2}} \,
\prod_{i=1}^{K^{\rm{II}}} \frac{{\sf v}_i - u_0 - \frac{\ell_0}{2}}{{\sf v}_i - u_0 - \frac{\ell_0-2k-2}{2}} \ \times
\label{symm tau1 T2} \\[1mm]
&\hspace{50mm} \prod_{i=1}^{K^{\rm{I}}} \[ \sqrt{\frac{x^-_i}{x^+_i}} \,
X_1^k \(u_0 - u_i \,, \ell_0 \,, \ell_i\)
\frac{x^{[\ell_0]}_0-x^+_i}{x^{[\ell_0]}_0-x^-_i} \] \Biggr\},
\notag \\[2mm]
\tau_{\ell_0,2} \[T_3 \,, T_2 \] &= S_{\vev{\ell_0-1,0}} \, \sum_{k=0}^{\ell_0-2} \Biggl\{
\prod_{i=1}^{K^{\rm{II}}} \frac{{\sf v}_i - u_0 - \frac{\ell_0}{2}}{{\sf v}_i - u_0 - \frac{\ell_0-2k-2}{2}} \ \times
\label{symm tau2} \\[1mm]
&\hspace{20mm} \prod_{i=1}^{K^{\rm{I}}} \[
\frac{x^-_i}{x^+_i} \,
\frac{1 - \frac{1}{x^{[\ell_0-2k-2]}_0 x^-_i}}{1 - \frac{1}{x^{[\ell_0-2k-2]}_0 x^+_i}} \,
X_1^{k+1} \(u_0 - u_i \,, \ell_0 \,, \ell_i\)
\frac{x^{[\ell_0]}_0-x^+_i}{x^{[\ell_0]}_0-x^-_i} \] \Biggr\}.
\notag
\end{alignat}
The functions $\tau_{\ell_0,0}$ and $\tau_{\ell_0,2} \[T_3 \,, T_2 \]$ can be combined together as
\begin{alignat}{3}
&\tau_{\ell_0,0} + \tau_{\ell_0,2} \left[ T_3 \,, T_2 \right] = S_{\vev{\ell_0-1,0}} \ \times
\label{symm tau02} \\[1mm]
&\quad \Biggl( 1 +
\prod_{i=1}^{K^{\rm{II}}}
\frac{{\sf v}_i - u_0 - \frac{\ell_0}{2}}{{\sf v}_i - u_0 + \frac{\ell_0}{2}} \,
\prod_{i=1}^{K^{\rm{I}}} \left[
\frac{x^{[-\ell_0]}_0 - x^-_i}{x^{[-\ell_0]}_0 - x^+_i} \,
X_0^{\ell_0} \( u_0 - u_i \,, \ell_0 \,, \ell_i \)
\frac{1 - \frac{1}{x^{[\ell_0]}_0 x^-_i}}{1 - \frac{1}{x^{[\ell_0]}_0 x^+_i}} \right]
\notag \\[1mm]
&\ \ \, + \sum_{k=1}^{\ell_0-1} \pare{
\prod_{i=1}^{K^{\rm{II}}}
\frac{{\sf v}_i - u_0 - \frac{\ell_0}{2}}{{\sf v}_i - u_0 - \frac{\ell_0-2k}{2}} \,
\prod_{i=1}^{K^{\rm{I}}} \left[
\frac{x^{[\ell_0-2k]}_0 - x^-_i}{x^{[\ell_0-2k]}_0 - x^+_i} \,
\, X_1^k \(u_0 - u_i \,, \ell_0 \,, \ell_i\)
\frac{x^{[\ell_0]}_0-x^+_i}{x^{[\ell_0]}_0-x^-_i} \right] }
\notag \\[1mm]
&\quad + \sum_{k=1}^{\ell_0-1} \Biggl\{
\prod_{i=1}^{K^{\rm{II}}} \frac{{\sf v}_i - u_0 - \frac{\ell_0}{2}}{{\sf v}_i - u_0 - \frac{\ell_0-2k}{2}} \,
\prod_{i=1}^{K^{\rm{I}}} \[
\frac{x^-_i}{x^+_i} \,
\frac{1 - \frac{1}{x^{[\ell_0-2k]}_0 x^-_i}}{1 - \frac{1}{x^{[\ell_0-2k]}_0 x^+_i}} \,
X_1^k \(u_0 - u_i \,, \ell_0 \,, \ell_i\)
\frac{x^{[\ell_0]}_0-x^+_i}{x^{[\ell_0]}_0-x^-_i} \] \Biggr\}
\Biggr) .
\notag
\end{alignat}
Let us compare the transfer matrix for the $\ell_0$\,-th symmetric
representation \eqref{eqn;FullEignvalue} with the conjectured one
\eqref{symmetric transfer 2}. Consider the fermionic terms first.
The fourth and fifth lines of \eqref{eqn;FullEignvalue} should be
compared with $\tau_{\ell_0,1} \[ T_3 \] + \tau_{\ell_0,1} \[ T_2
\]$ in \eqref{symm tau1 T3} and \eqref{symm tau1 T2}. From (\ref{xlolook})
one can deduce the identity\footnote{It should be noted that
$\mathcal{D}$ and $\mathscr{X}^{k,0}_k$ actually depend on $i$,
and that $\mathcal{D} = \mathscr{X}^{k,0}_k$ when $\ell_i=1$ for
each $i$.}
\begin{equation}
\frac{\mathscr{X}^{k,0}_k}{\mathcal{D}} \, \frac{u_0-u_i+\frac{\ell_0-\ell_i - 2k}{2}}{u_0-u_i+\frac{\ell_0-\ell_i}{2}} =
\prod_{j=1}^k \frac{u_0 - u_i + \frac{\ell_0 - \ell_i - 2j}{2}}{u_0 - u_i + \frac{\ell_0 + \ell_i - 2j}{2}} =
X_1^k \( u_0 - u_i \,, \ell_0 \,, \ell_i \).
\label{X1k identity}
\end{equation}
By using the relation
\begin{equation}
x_0^{[\pm \ell_0]} = x_0^\pm \,, \qquad
\mathcal{D} = \frac{x_0^- - x_i^+}{x_0^+ - x_i^-} \sqrt{\frac{x_0^+}{x_0^-} \, \frac{x_i^-}{x_i^+} } \, ,
\end{equation}
one finds nice agreement for the fermionic terms.

Next, let us look at the bosonic terms. The first, second, and third lines of \eqref{eqn;FullEignvalue} should be compared with $\tau_{\ell_0} + \tau_{\ell_0,2} \[ T_3,T_2 \]$ in \eqref{symm tau02}. The first term is $S_{\vev{\ell_0-1,0}}$ for both. The second term also agrees because of the relation
\begin{alignat}{3}
\frac{x^{[-\ell_0]}_0 - x^-_i}{x^{[-\ell_0]}_0 - x^+_i} \,
X_0^{\ell_0} &\( u_0 - u_i \,, \ell_0 \,, \ell_i \)
\frac{1 - \frac{1}{x^{[\ell_0]}_0 x^-_i}}{1 - \frac{1}{x^{[\ell_0]}_0 x^+_i}}
\notag \\[1mm]
&\quad = \frac{x^{[-\ell_0]}_0 - x^-_i}{x^{[\ell_0]}_0 - x^-_i} \, \frac{1 - \frac{1}{x^{[-\ell_0]}_0 x^+_i}}{1 - \frac{1}{x^{[\ell_0]}_0 x^+_i}} \,
\frac{u_0 - u_i + \frac{\ell_0 + \ell_i}{2}}{u_0 - u_i - \frac{\ell_0 + \ell_i}{2}} \, X_0^{\ell_0} \( u_0 - u_i \,, \ell_0 \,, \ell_i \),
\notag \\[1mm]
&\quad = \frac{x^{[-\ell_0]}_0 - x^-_i}{x^{[\ell_0]}_0 - x^-_i} \, \frac{1 - \frac{1}{x^{[-\ell_0]}_0 x^+_i}}{1 - \frac{1}{x^{[\ell_0]}_0 x^+_i}} \,
\frac{\mathscr{X}^{\ell_0,0}_{\ell_0}}{\mathcal{D}} \,.
\end{alignat}
Thus, we can proceed to identify the rest of the bosonic terms, namely the last two lines of the equation \eqref{symm tau02} and the third line of \eqref{eqn;FullEignvalue}. By substituting $x^{[\ell_0-2k]}$ of \eqref{x intermediate branches} into the definition of $\lambda_\pm (q,p_i,k)$ in \eqref{eqn;lambda-pm}, we find
\begin{equation}
\lambda_\pm (q,p_i,k) = \begin{cases}
\ds \ \frac{x_0^{[\ell_0-2k]} - x_i^-}{x_0^{[\ell_0-2k]} - x_i^+} \, \frac{u_0 - u_i + \frac{\ell_0 - \ell_i - 2k}{2}}{u_0 - u_i + \frac{\ell_0 - \ell_i}{2}} \,
\frac{x_0^{[\ell_0]} - x_i^+}{x_0^{[\ell_0]} - x_i^-} \,
\frac{\mathscr{X}^{k,0}_k}{\mathcal{D}} \,,
\\[6mm]
\ds \ \frac{x_i^-}{x_i^+} \, \frac{1 - \frac{1}{x_0^{[\ell_0-2k]} x_i^-}}{1 - \frac{1}{x_0^{[\ell_0-2k]} x_i^+}} \, \frac{u_0 - u_i + \frac{\ell_0 - \ell_i - 2k}{2}}{u_0 - u_i + \frac{\ell_0 - \ell_i}{2}} \,
\frac{x_0^{[\ell_0]} - x_i^+}{x_0^{[\ell_0]} - x_i^-} \,
\frac{\mathscr{X}^{k,0}_k}{\mathcal{D}} \,.
\end{cases}
\end{equation}
Thanks to the previous identity \eqref{X1k identity}, the rest of the bosonic terms \ad{of both} turn out to be identical. \ad{This completes the proof.}

For the reader's convenience, we rewrite here the transfer matrix for the $\ell_0$-th symmetric representation, obtained from the conjecture on the quantum characteristic function:
\begin{alignat}{3}
&T_{\vev{\ell_0-1,0}} (u_0|\{\vec u,\vec v,\vec w\}) = \prod_{i=1}^{K^{\rm{II}}} {\textstyle \frac{y_i-x^{[-\ell_0]}_0}{y_i - x^{[\ell_0]}_0}\sqrt{\frac{x^{[\ell_0]}_0}{x^{[-\ell_0]}_0}} } \ \times
\label{symmetric transfer 4} \\[2mm]
&\quad \Biggl( 1 +
\prod_{i=1}^{K^{\rm{II}}} {\textstyle
\frac{{\sf v}_i - u_0 - \frac{\ell_0}{2}}{{\sf v}_i - u_0 + \frac{\ell_0}{2}} } \,
\prod_{i=1}^{K^{\rm{I}}} {\textstyle \left[
\frac{x^{[-\ell_0]}_0 - x^-_i}{x^{[\ell_0]}_0 - x^-_i} \, \frac{1 - \frac{1}{x^{[-\ell_0]}_0 x^+_i}}{1 - \frac{1}{x^{[\ell_0]}_0 x^+_i}} \,
\frac{\mathscr{X}^{\ell_0,0}_{\ell_0}}{\mathcal{D}} \right] }
\notag \\[1mm]
&\quad + \sum_{k=1}^{\ell_0-1} \pare{
\prod_{i=1}^{K^{\rm{II}}} {\textstyle
\frac{{\sf v}_i - u_0 - \frac{\ell_0}{2}}{{\sf v}_i - u_0 - \frac{\ell_0-2k}{2}} } \,
\prod_{i=1}^{K^{\rm{I}}} {\textstyle \[
\frac{x^{[\ell_0-2k]}_0 - x^-_i}{x^{[\ell_0-2k]}_0 - x^+_i} \,
\frac{u_0-u_i+\frac{\ell_0-\ell_i - 2k}{2}}{u_0-u_i+\frac{\ell_0-\ell_i}{2}} \,
\frac{x^{[\ell_0]}_0-x^+_i}{x^{[\ell_0]}_0-x^-_i} \,
\frac{\mathscr{X}^{k,0}_k}{\mathcal{D}} \] }}
\notag \\[1mm]
&\quad + \sum_{k=1}^{\ell_0-1} \pare{
\prod_{i=1}^{K^{\rm{II}}} {\textstyle
\frac{{\sf v}_i - u_0 - \frac{\ell_0}{2}}{{\sf v}_i - u_0 - \frac{\ell_0-2k}{2}} }\,
\prod_{i=1}^{K^{\rm{I}}} {\textstyle \[
\frac{x^-_i}{x^+_i} \, \frac{1 - \frac{1}{x^{[\ell_0-2k]}_0 x^-_i}}{1 - \frac{1}{x^{[\ell_0-2k]}_0 x^+_i}} \,
\frac{u_0-u_i+\frac{\ell_0-\ell_i - 2k}{2}}{u_0-u_i+\frac{\ell_0-\ell_i}{2}} \,
\frac{x^{[\ell_0]}_0-x^+_i}{x^{[\ell_0]}_0-x^-_i} \,
\frac{\mathscr{X}^{k,0}_k}{\mathcal{D}} \] } }
\notag \\[1mm]
&\quad - \sum_{k=0}^{\ell_0-1} \pare{
\prod_{i=1}^{K^{\rm{III}}} {\textstyle
\frac{{\sf w}_i - u_0 - \frac{\ell_0-2k+1}{2}}{{\sf w}_i - u_0 - \frac{\ell_0-2k-1}{2}} } \,
\prod_{i=1}^{K^{\rm{II}}} {\textstyle
\frac{{\sf v}_i - u_0 - \frac{\ell_0}{2}}{{\sf v}_i - u_0 - \frac{\ell_0-2k}{2}} } \,
\prod_{i=1}^{K^{\rm{I}}} {\textstyle \[
\sqrt{\frac{x^-_i}{x^+_i}} \,
\frac{u_0-u_i+\frac{\ell_0-\ell_i - 2k}{2}}{u_0-u_i+\frac{\ell_0-\ell_i}{2}}  \,
\frac{x^{[\ell_0]}_0-x^+_i}{x^{[\ell_0]}_0-x^-_i}\frac{\mathscr{X}^{k,0}_k}{\mathcal{D}} \] }}
\notag \\[1mm]
&\quad - \sum_{k=0}^{\ell_0-1} \pare{
\prod_{i=1}^{K^{\rm{III}}} {\textstyle
\frac{{\sf w}_i - u_0 - \frac{\ell_0-2k-3}{2}}{{\sf w}_i - u_0 - \frac{\ell_0-2k-1}{2}} } \,
\prod_{i=1}^{K^{\rm{II}}} {\textstyle
\frac{{\sf v}_i - u_0 - \frac{\ell_0}{2}}{{\sf v}_i - u_0 - \frac{\ell_0-2k-2}{2}} } \,
\prod_{i=1}^{K^{\rm{I}}} {\textstyle
\[ \sqrt{\frac{x^-_i}{x^+_i}} \,
\frac{u_0-u_i+\frac{\ell_0-\ell_i - 2k}{2}}{u_0-u_i+\frac{\ell_0-\ell_i}{2}}  \,
\frac{x^{[\ell_0]}_0-x^+_i}{x^{[\ell_0]}_0-x^-_i}\frac{\mathscr{X}^{k,0}_k}{\mathcal{D}} \] }}
\Biggr) .
\notag
\end{alignat}
How the agreement works can be understood in the following way.
From the expression (\ref{symmetric transfer 4}) we see that, apparently, a spurious
dependence on the parameters $x_0^{[\ell_0 -2k]}$
is left as a
remnant of the fusion among the different blocks of the quantum
characteristic function. However, one can make use of (\ref{x
intermediate branches}) to re-express each of these variables only
in terms of the bound state variable $x_0^{[\ell_0]}$,
provided
one chooses a branch of the quadratic map. The remarkable
observation is that, after this replacement, one can recast the
above expression in a form that precisely agrees with our
result (\ref{eqn;FullEignvalue}). This happens for both choices of
branch, consistent with the fact that the formula we have obtained
{\it via} the alternative route of the ABA does not bear any
dependence on such a choice.

The transfer matrix for the symmetric representations \eqref{symmetric transfer 4} without auxiliary roots can be used
to compute the wrapping correction in the $\alg{sl}(2)$ sector
after the simultaneous flip $\( x_0^+ , x_i^+ \) \leftrightarrow
\( x_0^- , x_i^- \)$. By plugging the appropriate solution of the
asymptotic Bethe Ansatz equation into \eqref{symmetric transfer
4}, one can reproduce the results obtained by
\cite{Bajnok:2008qj,Gromov:2009tv}.

%%% Revision ends here %%%

\section*{Acknowledgements}

We thank Niklas Beisert, Matteo Beccaria, Sergey Frolov, Kolya
Gromov and Tomasz Lukowski for interesting discussions and helpful
correspondence, and the referees for providing several useful comments and remarks.
The work of G.~A. was supported in part by the
RFBI grant 08-01-00281-a, by the grant NSh-672.2006.1, by NWO
grant 047017015 and by the INTAS contract 03-51-6346. The work by
R.S. was supported by the Science Foundation Ireland under Grant
No. 07/RFP/PHYF104.

\appendix

\section{Elements of the bound state scattering matrix}\label{A}

In this appendix we list the elements of the bound state S-matrix
from \cite{Arutyunov:2009mi} that are used in this paper. We start
with the Case I S-matrix coefficients\footnote{We suppress the dependence on momenta in order to have a lighter notation.
All functions appearing in this section have to be understood as $\mathscr{X} \equiv \mathscr{X}(p_1, p_2)$, $\mathscr{Y} \equiv \mathscr{Y}(p_1, p_2)$, $\mathscr{Z} \equiv \mathscr{Z}(p_1, p_2)$ (indices are omitted here for simplicity) and  $\mathcal{D}=\mathcal{D}(p_1, p_2)$.}
\begin{eqnarray}
\mathscr{X}^{k,l}_n &=&(-1)^{k+n} \, \pi \mathcal{D} \frac{\sin
[(k-\ell_1) \pi ] \, \Gamma (l+1)}{\sin [\ell_1 \pi] \sin [(k +l
-\ell_2-n) \pi ] \, \Gamma (l-\ell_2+1) \Gamma
   (n+1)} \times \nonumber\\
&& \frac{\Gamma
   (n+1-\ell_1) \Gamma
   \left(l+\frac{\ell_1-\ell_2}{2}-n-\delta u
   \right) \Gamma \left(1-\frac{\ell_1+\ell_2}{2}-\delta u \right)}{\Gamma
   \left(k+l-\frac{\ell_1+\ell_2}{2}-\delta u +1\right) \Gamma \left(\frac{\ell_1-\ell_2}{2}- \delta u \right)} \times \\
&& _4\tilde{F}_3
   \left(-k,-n,\delta u
   +1-\frac{\ell_1-\ell_2}{2} ,\frac{\ell_2-\ell_1}{2}-\delta u; 1-\ell_1,\ell_2-k-l,l-n+1;1 \right).\nonumber
\end{eqnarray}
One has defined $_4\tilde{F}_3 (x,y,z,t;r,v,w;\tau) = {_4F_3}
(x,y,z,t;r,v,w;\tau)/[\Gamma (r) \Gamma (v) \Gamma (w)]$,
\begin{eqnarray}
\mathcal{D}= \frac{x_1^- -x_2^+}{x_1^+
-x_2^-}\frac{e^{i\frac{p_1}{2}}}{e^{i\frac{p_2}{2}}}
\end{eqnarray}
and
$$
\delta u = u_1 - u_2.
$$
The relevant entries of the Case II S-matrix are given by
\begin{eqnarray}
\mathscr{Y}^{k,0;1}_{k;1} &=&
\frac{x^+_1-x^+_2}{x^-_1-x^+_2}\sqrt{\frac{x^-_1}{x^+_1}}\left[1-\frac{k}{\delta
u+\frac{\ell_1-\ell_2}{2} }\right]\mathscr{X}^{k,0}_{k},\\
\mathscr{Y}^{k,0;2}_{k;2}
&=&\frac{x^-_1-x^-_2}{x^-_1-x^+_2}\sqrt{\frac{x^+_2}{x^-_2}}\mathscr{X}^{k,0}_{k},\\
\mathscr{Y}^{k,0;1}_{k;2}
&=&\frac{x^-_2-x^+_2}{x^-_1-x^+_2}\sqrt{\frac{x^-_1 x^+_2}{x^+_1
x^-_2}}\frac{\sqrt{\ell_1}\eta(p_1)}{\sqrt{\ell_2}\eta(p_2)}\frac{k-\ell_1}{\ell_1}\mathscr{X}^{k,0}_{k},\\
\mathscr{Y}^{k,0;2}_{k;1}
&=&\frac{x^+_1-x^-_1}{x^-_1-x^+_2}\frac{\sqrt{\ell_2}\eta(p_2)}{\sqrt{\ell_1}\eta(p_1)}\mathscr{X}^{k,0}_{k},\\
\mathscr{Y}^{k,0;4}_{k;1} &=&\frac{\sqrt{\ell_1\ell_2}\eta(p_1)\eta(p_2)}{x^+_1 x^+_2-1}\frac{k}{i\ell_1}\mathscr{X}^{k,0}_{k},\\
\mathscr{Y}^{k,0;4}_{k;4}
&=&\sqrt{\frac{x^+_2}{x^-_2}}\frac{x^+_1x^-_2-1}{x^+_1x^+_2-1}\mathscr{X}^{k,0}_{k},\\
\mathscr{Y}^{k,0;1}_{k;4} &=&
\frac{i}{\sqrt{\ell_1\ell_2}\eta(p_1)\eta(p_2)}\sqrt{\frac{x^+_1x^+_2}{x^-_1x^-_2}}\frac{(x^-_1-x^+_1)(x^-_2-x^+_2)}{x^+_1x^+_2-1}\mathscr{X}^{k,0}_{k},
\end{eqnarray}
Finally, from the Case III S-matrix we used
\begin{eqnarray}
\mathscr{Z}^{k,0;1}_{k;1} &=&\left[1-\frac{2ik}{g}\frac{x^+_1
(x^-_2-x^-_1 x^+_1
x^+_2)}{(x^-_2-x^+_1)(1-x^-_1x^+_1)(1-x^+_1x^+_2)}\right]\frac{\mathscr{X}^{k,0}_{k}}{\mathcal{D}},\\
\mathscr{Z}^{k,0;1}_{k;3} &=& \frac{2
(x^-_1-x^+_1)(x^+_1)^2(x^-_2-x^+_2)} {g
(x^+_1-x^-_2)(1-x^+_1x^-_1) (1-x^+_1
x^+_2)\eta(p_1)^2}\frac{\mathscr{X}^{k,0}_{k}}{\mathcal{D}},\\
\mathscr{Z}^{k,0;3}_{k;1} &=& \frac{i k (\ell_1-k)}{\ell_1}\frac{
(x^-_2-x^+_2) \eta(p_1)^2}{(x^+_1-x^-_2)(1-x^+_1
x^+_2)}\frac{\mathscr{X}^{k,0}_{k}}{\mathcal{D}},\\
\mathscr{Z}^{k,0;3}_{k;3} &=&\left[
\frac{(x^+_1-x^+_2)(1-x^-_2x^+_1)}{(x^+_1-x^-_2)(1-x^+_1 x^+_2)} +
\frac{2ik}{g}\frac{x^+_1
(x^+_2-x^-_1x^+_1x^-_2)}{(x^+_1-x^-_2)(1-x^-_1
x^+_1)(1-x^+_1x^+_2)}\right]\frac{\mathscr{X}^{k,0}_{k}}{\mathcal{D}}\quad
\quad
\end{eqnarray}
and
\begin{eqnarray}
\mathscr{Z}^{k,0;6}_{k;1} &=&
\frac{ik\sqrt{\ell_2}\eta(p_1)\eta(p_2)}{\sqrt{\ell_1}}\frac{
x^-_1-x^-_2 }{(x^-_1-x^+_2)
(1-x^+_1x^+_2)}\sqrt{\frac{x^+_2}{x^-_2}}\mathscr{X}^{k,0}_k,\\
\mathscr{Z}^{k,0;6}_{k;3}
&=&\sqrt{\frac{\ell_2}{\ell_1}}\frac{\eta(p_2)}{\eta(p_1)} \frac{
(x^-_1-x^+_1)(x^-_2
x^+_1-1)}{(x^-_1-x^+_2)(x^+_1x^+_2-1)}\sqrt{\frac{x^+_2}{x^-_2}}
\mathscr{X}^{k,0}_k,\\
\mathscr{Z}^{k,0;6}_{k;6}&=&\frac{(x^-_1-x^-_2)(x^-_2x^+_1-1)x^+_2}
{(x^-_1-x^+_2)(x^+_1 x^+_2-1)x^-_2}\mathscr{X}^{k,0}_k,\\
\mathscr{Z}^{k,0;3}_{k;6}&=&\frac{(\ell_1-k)\eta
(p_1)}{\sqrt{\ell_1\ell_2}\eta
(p_2)}\frac{(x^-_2x^+_1-1)(x^-_2-x^+_2) x^+_2}{
(x^-_1-x^+_2)(x^+_1x^+_2-1)x^-_2}\sqrt{\frac{x^-_1}{x^+_1}}\mathscr{X}^{k,0}_k,\\
\mathscr{Z}^{k,0;1}_{k;6}&=&\frac{i}{\sqrt{\ell_1\ell_2}\eta(p_1)
\eta(p_2)}\frac{(x^-_1-x^-_2) (x^-_1-x^+_1)(x^-_2-x^+_2) x^+_2}{
(x^-_1-x^+_2)(x^+_1 x^+_2-1)x^-_2}
\sqrt{\frac{x^+_1}{x^-_1}}\mathscr{X}^{k,0}_k.
\end{eqnarray}

\section{Algebraic Bethe ansatz for the 6-vertex
model}\label{sect;6Vertex}

In the this paper we use the algebraic Bethe ansatz approach to
diagonalize the $\ads$ superstring transfer matrix for bound
states. We will closely follow the discussion for the Hubbard
model \cite{martins-1997}. In this model, just as in our case, the
6-vertex model plays an important role. In this section we will
discuss the algebraic Bethe ansatz for this model for completeness
and to fix notations.

The algebraic Bethe ansatz for the 6-vertex model is a standard
chapter of the theory of integrable systems, and it is treated for
example in \cite{Faddeev:1996iy,Korepin}. The scattering matrix of
the model is given by
\begin{eqnarray}
r_{12}(u_1,u_2) = \begin{pmatrix}
  1 & 0 & 0 & 0 \\
  0 & b(u_1,u_2) & a(u_1,u_2) & 0 \\
  0 & a(u_1,u_2) & b(u_1,u_2) & 0 \\
  0 & 0 & 0 & 1
\end{pmatrix},
\end{eqnarray}
where
\begin{eqnarray}
a = \frac{U}{u_1-u_2+U}, \qquad b = \frac{u_1-u_2}{u_1-u_2+U}.
\end{eqnarray}
It is convenient to write it as
\begin{eqnarray}\nonumber
r_{12}(u_1,u_2)
&=& r_{\alpha\beta}^{\gamma\delta}(u_1,u_2)E^{\alpha}_{\gamma}\otimes E^{\beta}_{\delta}\\
&=& \frac{u_1-u_2}{u_1-u_2+U}\left[E^\alpha_\alpha\otimes
E^\beta_\beta +\frac{U}{u_1-u_2} E^\alpha_\beta\otimes
E^\beta_\alpha\right]\label{eqn;6vertexComponents},
\end{eqnarray}
with $E^\alpha_\beta$ the standard matrix unities. Let us consider
$K$ particles, with rapidities $u_i$. Now, one can construct the
monodromy matrix
\begin{eqnarray}
\mathcal{T}(u_0|\vec{u})= \prod_{i=1}^{K} r_{0i}(u_0|u_i).
\end{eqnarray}
Let us write it as a matrix in the auxiliary space
\begin{eqnarray}
\mathcal{T}^{(1)}(u_0|\vec{u})= \begin{pmatrix}
  A(u_0|\vec{u}) & B(u_0|\vec{u}) \\
  C(u_0|\vec{u}) & D(u_0|\vec{u})
\end{pmatrix}.
\end{eqnarray}
In the algebraic Bethe Ansatz, one constructs the eigenvalues of
the transfer matrix by first specifying a ground state
$|0\rangle$. The ground state, in this case, is defined as
\begin{eqnarray}
|0\rangle = \bigotimes_{i=1}^{K}\begin{pmatrix}
  ~1  \\
  ~0 &
\end{pmatrix}.
\end{eqnarray}
It is easily checked that it is an eigenstate of the transfer
matrix. To be a bit more precise, the action of the different
elements of the monodromy matrix on $|0\rangle$ is given by
\begin{eqnarray}
 A(u_0|\vec{u})|0\rangle &=& |0\rangle,\nonumber\\
 C(u_0|\vec{u})|0\rangle &=& 0,\\
 D(u_0|\vec{u})|0\rangle &=& \prod_{i=1}^{K}b(u_0,u_i)|0\rangle\nonumber.
\end{eqnarray}
Thus, $|0\rangle$ is an eigenstate of the transfer matrix with the
following eigenvalue
\begin{eqnarray}
1+\prod_{i=1}^{K}b(u_0,u_i).
\end{eqnarray}
The field $B$ will be considered as a creation operator. It will
create all the other eigenstates out of the vacuum. We introduce
additional parameters $w_i$ and consider the state
\begin{eqnarray}
|M\rangle := \phi_M(w_1,\ldots,w_M)|0\rangle,\qquad
\phi_M(w_1,\ldots,w_M):=\prod_{i=1}^{M} B(w_i|\vec{u}).
\end{eqnarray}
In the context of the Heisenberg spin chain the vacuum corresponds
to all spins down and the state $|M\rangle$ corresponds to the
eigenstate of the transfer matrix that has $M$ spins turned up.

In order to evaluate the action of the transfer matrix
$\mathscr{T}(u_0|\vec{u}) = A(u_0|\vec{u})+D(u_0|\vec{u})$ on the
state $|M\rangle$, one needs the commutation relations between the
fields $A,B,D$. From (\ref{eqn;YBE-operators}) one reads
\begin{eqnarray}
A(u_0|\vec{u})B(w|\vec{u})&=&\frac{1}{b(w,u_0)}B(w|\vec{u})A(u_0|\vec{u})
-\frac{a(w,u_0)}{b(w,u_0)}B(u_0|\vec{u})A(w|\vec{u})\nonumber\\
B(w_1|\vec{u})B(w_2|\vec{u})&=&B(w_2|\vec{u})B(w_1|\vec{u})\\
D(u_0|\vec{u})B(w|\vec{u})&=&\frac{1}{b(u_0,w)}B(w|\vec{u})D(u_0|\vec{u})
-\frac{a(u_0,w)}{b(u_0,w)}B(u_0|\vec{u})D(w|\vec{u}).\nonumber
\end{eqnarray}
From this, one can compute exactly when $|M\rangle$ is an
eigenstate of the transfer matrix. By definition we have that
\begin{eqnarray}\label{eqn;XXXinduction}
|M\rangle = B(w_M|\vec{u})|M-1\rangle,
\end{eqnarray}
and this allows us to use induction. By using the identity
\begin{eqnarray}
\frac{1}{b(w_M,u_0)}\frac{a(w_i,u_0)}{b(w_i,u_0)}-\frac{a(w_M,u_0)}{b(w_Mu_0)}\frac{a(w_i,w_M)}{b(w_i,w_M)}
= \frac{a(w_i,u_0)}{b(w_i,u_0)}\frac{1}{b(w_M,w_i)}
\end{eqnarray}
in (\ref{eqn;XXXinduction}) one can prove
\begin{eqnarray}
A(u_0|\vec{u})\phi_M(w_1,\ldots,w_M) &=& \prod_{i=1}^{M}\frac{1}{b(w_i,u_0)}\phi_M(w_1,\ldots,w_M)A(u_0|\vec{u})\\
&&-
\sum_{i=1}^{M}\left[\frac{a(w_i,u_0)}{b(w_i,u_0)}\prod_{j=1,j\neq
i}^{M}\frac{1}{b(w_j,w_i)}\hat{\phi}_M
A(w_i|\vec{u})\right],\nonumber
\end{eqnarray}
where $\hat{\phi}_M $ stands for
$\phi_M(\ldots,w_{i-1},u_0,w_{i+1},\ldots)$. One can find a
similar relation for the commutator between $D$ and $B$. By using
these relations one finds that
\begin{eqnarray}
\mathscr{T}(u_0|\vec{u})|M\rangle &=&
\left\{A(u_0|\vec{u})+D(u_0|\vec{u})\right\}|M\rangle\\
&=&\phi_M(w_1,\ldots,w_M)\left\{A(u_0|\vec{u})\prod_{i=1}^{M}\frac{1}{b(w_i,u_0)}
+D(u_0|\vec{u})\prod_{i=1}^{M}\frac{1}{b(u_0,w_i)}\right\}|0\rangle \nonumber\\
&&-\sum_{i=1}^{M}\left[\frac{a(w_i,u_0)}{b(w_i,u_0)}\, \hat{\phi}_M \left\{\prod_{j\neq
i}\frac{1}{b(w_j,w_i)} A(w_i|\vec{u})-\prod_{j\neq
i}\frac{1}{b(w_i,w_j)}
D(w_i|\vec{u})\right\}\right]|0\rangle.\nonumber
\end{eqnarray}
From this we find that $|M\rangle$ is an eigenstate of the
transfer matrix with eigenvalue
\begin{eqnarray}\label{eqn;eigenvalue6V}
\Lambda^{(6v)}(u_0|\vec{u}) = \prod_{i=1}^M
\frac{1}{b(w_i,u_0)}+\prod_{i=1}^M
\frac{1}{b(u_0,w_i)}\prod_{i=1}^K b(u_0,u_i)
\end{eqnarray}
provided that the auxiliary parameters $w_i$ satisfy the following
equations
\begin{eqnarray}\label{eqn;AuxEqns6V}
\prod_{i=1}^K b(w_j,u_i)=\prod_{i=1,i\neq j}^M
\frac{b(w_j,w_i)}{b(w_i,w_j)}.
\end{eqnarray}
This now completely determines the spectrum of the 6-vertex model.

To conclude, we briefly explain how these eigenvalues are used to
generate an infinite tower of conserved charges. From
(\ref{eqn;YBE-operators}) one finds that
\begin{eqnarray}
\mathscr{T}(u_0|\vec{u})\mathscr{T}(\mu|\vec{u})=\mathscr{T}(\mu|\vec{u})\mathscr{T}(u_0|\vec{u}).
\end{eqnarray}
This means that if one writes $\mathscr{T}(u_0|\vec{u})$ as a
series the auxiliary parameter $u_0$, the coefficients of this
series will depend on $\vec{u}$ and are in involution with each
other. It actually turns out that the 6-vertex model Hamiltonian
can be written in terms of these coefficients and hence this means
that the coefficients of $\mathscr{T}(u_0|\vec{u})$ yield an
infinite number of conserved quantities.

\section{Alternative vacua}\label{sec;TasBA}

In this section we will discuss a class of higher excited states
for which we present a full derivation of its eigenvalue of the
transfer matrix and the auxiliary Bethe equations. From the
general construction it is easily seen that a more general
eigenvector of the transfer matrix is given by
\begin{eqnarray}
|a\rangle = \Phi(\lambda_1,\ldots,\lambda_a)|0\rangle_P, \qquad
\Phi(\lambda_1,\ldots,\lambda_a) = B_3(\lambda_1)\ldots
B_3(\lambda_a).
\end{eqnarray}
These states have quantum number $K^{\rm{III}}=0$. This allows for
a similar inductive procedure as applied to the 6-vertex model in
section \ref{sect;6Vertex}. Furthermore, because of the properties
of the creation operators (\ref{eqn;CommRelCreation}), we find
that
\begin{eqnarray}\label{eqn;symmetryTas}
\Phi(\lambda_1,\ldots,\lambda_{j-1},\lambda_{j},\ldots\lambda_a) =
-\mathscr{X}^{0,0}_0
r^{33}_{33}(\lambda_{j-1},\lambda_{j})\Phi(\lambda_1,\ldots,\lambda_{j},\lambda_{j-1},\ldots\lambda_a).
\end{eqnarray}
This means that all permutations of the momenta $\lambda_i$ are
related to each other by a simple multiplication by a scalar
prefactor. We will exploit this property later on. Let us first
derive some useful identities. One uses induction to show that
\begin{eqnarray}
A^{4}_3|a\rangle = C^*_3|a\rangle = C_4|a\rangle = C|a\rangle=0,
\end{eqnarray}
for any $a$. This vastly simplifies the computations, since we can
discard any term proportional to the above operators from the
commutation relations. Let us first turn to (\ref{eqn;CommRel}).
This now becomes, after discarding the term proportional to
$C^*_3$,
\begin{eqnarray}
\mathcal{T}_{3,k}^{3,k}(q)B_{3}(\lambda)
=\frac{\mathscr{X}^{k,0}_k}{\mathscr{Y}^{k,0;1}_{k;1}}B_{3}(\lambda)\mathcal{T}_{3,k}^{3,k}(q)
+\frac{\mathscr{Y}^{k,1;1}_{k;2}}{\mathscr{Y}^{k,0;1}_{k;1}}\mathcal{T}_{3,k}^{k}(q)A_{3}^{3}(\lambda)
+\frac{\mathscr{Y}^{k,1;1}_{k;4}}{\mathscr{Y}^{k,0;1}_{k;1}}\mathcal{T}_{3,k}^{34,k-1}(q)A_{3}^{3}(\lambda).\nonumber
\end{eqnarray}
Applying this to $\Phi(\lambda_1,\ldots,\lambda_a) =
B_3(\lambda_1)\Phi(\lambda_2,\ldots,\lambda_a)$ we find
\begin{eqnarray}\label{eqn;TasComm1}
\mathcal{T}_{3,k}^{3,k}(q)\Phi(\lambda_1,\ldots,\lambda_a)&=&
\frac{\mathscr{X}^{k,0}_k}{\mathscr{Y}^{k,0;1}_{k;1}}B_{3}(\lambda_1)\mathcal{T}_{3,k}^{3,k}(q)\Phi(\lambda_2,\ldots,\lambda_a)\nonumber\\
&&+\frac{\mathscr{Y}^{k,1;1}_{k;2}}{\mathscr{Y}^{k,0;1}_{k;1}}\mathcal{T}_{3,k}^{k}(q)A_{3}^{3}(\lambda_1)\Phi(\lambda_2,\ldots,\lambda_a)\\
&&+\frac{\mathscr{Y}^{k,1;1}_{k;4}}{\mathscr{Y}^{k,0;1}_{k;1}}\mathcal{T}_{3,k}^{34,k-1}(q)A_{3}^{3}(\lambda_1)\Phi(\lambda_2,\ldots,\lambda_a).\nonumber
\end{eqnarray}
Obviously, by applying this relation recursively one finds
\begin{eqnarray}
\mathcal{T}_{3,k}^{3,k}(q)\Phi(\lambda_1,\ldots,\lambda_a)&=&\prod_{i=1}^{a}\frac{\mathscr{X}^{k,0}_k(q,\lambda_i)}{\mathscr{Y}^{k,0;1}_{k;1}(q,\lambda_i)}\Phi(\lambda_1,\ldots,\lambda_a)\mathcal{T}_{3,k}^{3,k}(q)\nonumber\\
&&+\sum_{i=1}^{a}c_i\Phi_{k;i}(q,\lambda)A_{3}^{3}(\lambda_i)+\sum_{i=1}^{a}d_i\Psi_{k;i}(q,\lambda)A_{3}^{3}(\lambda_i),\qquad
\end{eqnarray}
where $c_i$ are some numerical coefficients and
$\Phi_{k;i}(q,\lambda) = \mathcal{T}_{3,k}^{k}(q) \prod_{j\neq i}
B_3(\lambda_i), \Psi_{k;i}(q,\lambda) =
\mathcal{T}_{3,k}^{34,k-1}(q) \prod_{j\neq i} B_3(\lambda_i)$. It
is easily seen from (\ref{eqn;TasComm1}) that the numerical
coefficients in front of
$\Phi_{k;1}(q,\lambda),\Psi_{k;1}(q,\lambda)$ are given by
\begin{eqnarray}
c_1 =
\frac{\mathscr{Y}^{k,1;1}_{k;2}(q,\lambda_1)}{\mathscr{Y}^{k,0;1}_{k;1}(q,\lambda_1)}
\prod_{i=2}^{a}\frac{\mathscr{X}^{k,0}_k(q,\lambda_i)}{\mathscr{Y}^{k,0;1}_{k;1}(q,\lambda_i)},
\qquad d_1 =
\frac{\mathscr{Y}^{k,1;1}_{k;4}(q,\lambda_1)}{\mathscr{Y}^{k,0;1}_{k;1}(q,\lambda_1)}
\prod_{i=2}^{a}\frac{\mathscr{X}^{k,0}_k(q,\lambda_i)}{\mathscr{Y}^{k,0;1}_{k;1}(q,\lambda_i)}.
\end{eqnarray}
Here we can exploit the symmetry property (\ref{eqn;symmetryTas})
to relate all the other coefficients to this one. Let us denote
these proportionality coefficients by $\mathcal{P}_{1i}$, then we
find
\begin{eqnarray}
\mathcal{T}_{3,k}^{3,k}(q)\Phi(\lambda_1,\ldots,\lambda_a)&=&\prod_{i=1}^{a}\frac{\mathscr{X}^{k,0}_k(q,\lambda_i)}{\mathscr{Y}^{k,0;1}_{k;1}(q,\lambda_i)}\Phi(\lambda_1,\ldots,\lambda_a)\mathcal{T}_{3,k}^{3,k}(q)\nonumber\\
&&+\sum_{i=1}^{a}c_iP_{1i}\Phi_{k;i}(q,\lambda)A_{3}^{3}(\lambda_i)+\sum_{i=1}^{a}d_iP_{1i}\Psi_{k;i}(q,\lambda)A_{3}^{3}(\lambda_i),\qquad~
\end{eqnarray}
where
\begin{eqnarray}
c_j =
\frac{\mathscr{Y}^{k,1;1}_{k;2}(q,\lambda_j)}{\mathscr{Y}^{k,0;1}_{k;1}(q,\lambda_j)}
\prod_{i=1,i\neq j
}^{a}\frac{\mathscr{X}^{k,0}_k(q,\lambda_i)}{\mathscr{Y}^{k,0;1}_{k;1}(q,\lambda_i)},\qquad
d_j =
\frac{\mathscr{Y}^{k,1;1}_{k;4}(q,\lambda_j)}{\mathscr{Y}^{k,0;1}_{k;1}(q,\lambda_j)}
\prod_{i=1,i\neq j
}^{a}\frac{\mathscr{X}^{k,0}_k(q,\lambda_i)}{\mathscr{Y}^{k,0;1}_{k;1}(q,\lambda_i)}.~~
\end{eqnarray}
Next, we consider the commutator with $\mathcal{T}^k_k +
\mathcal{T}^{34,k-1}_{34,k-1}$. Upon dismissing vanishing
terms we find
\begin{eqnarray}
\left[\mathcal{T}^k_k +
\mathcal{T}^{34,k-1}_{34,k-1}\right]B_3(\lambda) &=&
\frac{\mathscr{X}^{k,0}_k}{\mathscr{Y}^{k,0;1}_{k;1}}B_3(\lambda)\left[\mathcal{T}^k_k
+ \mathcal{T}^{34,k-1}_{34,k-1}\right] +\\
&&\frac{\mathscr{Y}^{k,1;1}_{k;2}}{\mathscr{Y}^{k,0;1}_{k;1}}\left\{\mathcal{T}^{k}_{3,k}B-\mathcal{T}^{4,k-1}_{34,k-1}A^3_3\right\}
+\frac{\mathscr{Y}^{k,1;1}_{k;4}}{\mathscr{Y}^{k,0;1}_{k;1}}\left\{\mathcal{T}^{34,k-1}_{3,k}B+\mathcal{T}^{4,k-1}_{k}A^3_3\right\}.\nonumber
\end{eqnarray}
If we now define $\hat{\Phi}_{k;i}(q,\lambda) =
\mathcal{T}^{4,k-1}_{k}(q) \prod_{j\neq i}
\ad{B_3(\lambda_j)},\hat{\Psi}_{k;i}(q,\lambda) =
\mathcal{T}^{4,k-1}_{34,k-1}(q) \prod_{j\neq i} \ad{B_3(\lambda_j)}$,
then we can repeat the above steps to find
\begin{eqnarray}
\left[\mathcal{T}^k_k +
\mathcal{T}^{34,k-1}_{34,k-1}\right]\Phi(\lambda_1,\ldots,\lambda_a)
&=&
\prod_{i=1}^{a}\frac{\mathscr{X}^{k,0}_k(q,\lambda_i)}{\mathscr{Y}^{k,0;1}_{k;1}(q,\lambda_i)}\Phi(\lambda_1,\ldots,\lambda_a)\left[\mathcal{T}^k_k
+ \mathcal{T}^{34,k-1}_{34,k-1}\right] +\nonumber\\
&&\sum_{i=1}^{a}
c_iP_{1i}\left\{\Phi_{k;i}(q,\lambda)B(\lambda_i)-\hat{\Psi}_{k;i}(q,\lambda)A^3_3(\lambda_i)\right\}
+\\
&&\sum_{i=1}^{a}
d_iP_{1i}\left\{\Psi_{k;i}(q,\lambda)B(\lambda_i)+\hat{\Phi}_{k;i}(q,\lambda)A^3_3(\lambda_i)\right\}.\qquad\nonumber
\end{eqnarray}
The last commutation relation finally gives
\begin{eqnarray}
\mathcal{T}_{4,k}^{4,k}(q)\Phi(\lambda_1,\ldots,\lambda_a)&=&
\frac{\mathscr{X}^{k+1,0}_{k+1}}{\mathscr{Y}^{k,0;1}_{k;1}}\frac{u_q-u_{\lambda_1}+\frac{\ell_0-1}{2}-k}{u_q-u_{\lambda_1}+\frac{\ell_0-3}{2}-k}B_{3}(\lambda_1)\mathcal{T}_{4,k}^{4,k}(q)\Phi(\lambda_2,\ldots,\lambda_a)\nonumber\\
&&-\frac{\mathscr{Y}^{k+1,1;1}_{k+1;2}}{\mathscr{Y}^{k+1,0;1}_{k+1;1}}\mathcal{T}_{34,k}^{4,k}(q)B(\lambda_1)\Phi(\lambda_2,\ldots,\lambda_a)\\
&&+\frac{\mathscr{Y}^{k+1,1;1}_{k+1;4}}{\mathscr{Y}^{k+1,0;1}_{k+1;1}}\mathcal{T}_{k+1}^{4,k}(q)B(\lambda_1)\Phi(\lambda_2,\ldots,\lambda_a).\nonumber
\end{eqnarray}
Again the same arguments apply as above.

By summing now all the terms, we find that $|a\rangle$ is indeed
an eigenstate of the transfer matrix, provided that the parameters
$\lambda_i$ satisfy
\begin{eqnarray}
B(\lambda_i)|0\rangle_P = A^3_3(\lambda_i)|0\rangle_P.
\end{eqnarray}
When working this out, we see that the above only depends on
$x^+(\lambda_i)$, which we denote as $y_i\equiv x^+(\lambda_i)$.
The explicit formula is given by
\begin{eqnarray}
\prod_{j=1}^{K^{\rm{I}}}
\frac{y_i-x^+_j}{y_i-x^-_j}\sqrt{\frac{x_j^-}{x^+_j}}= 1,
\end{eqnarray}
which agrees with the known auxiliary BAE. The explicit eigenvalue
of $|a\rangle$ is given by
\begin{eqnarray}
\Lambda(q|\vec{p}) &=&
\prod_{m=1}^{K^{\rm{II}}}\frac{\mathscr{X}^{0,0}_0(q,\lambda_m)}{\mathscr{Y}^{0,0;1}_{0;1}(q,\lambda_m)}
+ \prod_{i=1}^{K^{\rm{I}}} \mathscr{Z}^{\ell_0,0;1}_{\ell_0;1}(q,p_i)\prod_{m=1}^{K^{\rm{II}}}\frac{\mathscr{X}^{\ell_0,0}_{\ell_0}(q,\lambda_m)}{\mathscr{Y}^{\ell_0,0;1}_{\ell_0;1}(q,\lambda_m)}+\nonumber\\
&&\sum_{k=1}^{\ell_0-1}
\prod_{m=1}^{K^{\rm{II}}}\frac{\mathscr{X}^{k,0}_k(q,\lambda_m)}{\mathscr{Y}^{k,0;1}_{k;1}(q,\lambda_m)}\left\{\prod_{i=1}^{K^{\rm{I}}}\lambda_+(q,p_i)
+\prod_{i=1}^{K^{\rm{I}}}\lambda_-(q,p_i)\right\}+\\
&&-\sum_{k=0}^{\ell_0-1}
\prod_{m=1}^{K^{\rm{II}}}\frac{\mathscr{X}^{k,0}_k(q,\lambda_m)}{\mathscr{Y}^{k,0;1}_{k;1}(q,\lambda_m)}
\left[1+\frac{u_q-u_{\lambda_m}+\frac{\ell_0}{2}-k}{u_q-u_{\lambda_m}+\frac{\ell_0-2}{2}-k}\right]\prod_{i=1}^{K^{\rm{I}}}\mathscr{Y}^{k,0;1}_{k;1}(q,p_i).\nonumber
\end{eqnarray}
This is indeed the case $K^{\rm{III}}=0$ of (\ref{eqn;FullEignvalue}).

\bibliographystyle{JHEP}
\bibliography{LitRmat}

\end{document}